\documentclass[12pt,a4paper,twoside]{article}
\usepackage{graphicx,cite}

\def\be{\begin{equation}}
\def\ee{\end{equation}}
\def\bea{\begin{eqnarray}}
\def\eea{\end{eqnarray}}
\def\nnb{\nonumber}
\def\dps{\displaystyle}
\def\lk{\left(}
\def\rk{\right)}
\def\lek{\left[}
\def\rek{\right]}
\def\bbuildrel#1_#2^#3{\mathrel{\mathop{\kern 0pt#1}\limits_{#2}^{#3}}}
\def\slash#1{\setbox0=\hbox{$#1$}#1\hskip-\wd0\dimen0=5pt\advance
       \dimen0 by-\ht0\advance\dimen0 by\dp0\lower0.5\dimen0\hbox
         to\wd0{\hss\sl/\/\hss}}
\def\gev{{\rm GeV}}
\def\mev{{\rm MeV}}

\newcommand{\gae}{\lower 2pt \hbox{$\, \buildrel {\scriptstyle >}\over
    {\scriptstyle \sim}\,$}}
\newcommand{\lae}{\lower 2pt \hbox{$\, \buildrel {\scriptstyle <}\over
    {\scriptstyle \sim}\,$}}

\newcommand{\f}{\frac}
\newcommand{\fm}[2]{{\textstyle \frac{#1}{#2}}}
\newcommand{\me}[1]{\langle#1\rangle}

\newcommand{\as}{\widetilde{\alpha}_{\mathrm s}}

\newcommand{\s}{\hat{s}}

\newcommand{\spp}{\vphantom{\bigg(}}

\begin{document}

\begin{titlepage}

\begin{flushright}
CERN-PH-TH/2007-177\\
SLAC-PUB-12859\\
ZU-TH 30/07\\
FERMILAB-PUB-07-639-T\\
PITHA 07/21\\[2cm]
\end{flushright}

\begin{center}
\setlength {\baselineskip}{0.3in} {\bf\Large 
Logarithmically Enhanced Corrections to the \\Decay Rate and Forward Backward Asymmetry \\in $\bar{B} \to X_s \ell^+ \ell^-$}\\[17mm]

\setlength {\baselineskip}{0.2in} {\large\bf  Tobias Huber$^{1,2}$, 
Tobias Hurth$^{3,4,}$\footnote{Heisenberg Fellow}, and Enrico Lunghi$^5$}\\[5mm]

$^1$~{\it Institute for Theoretical Physics, Univ.  of Zurich, CH-8057, Zurich, Switzerland.}\\[3mm]	  

$^2$~{\it Institut f\"ur Theoretische Physik E, RWTH Aachen, D-52056 Aachen, Germany}\\[3mm]
	  
$^3$~{\it CERN, Dept. of Physics, Theory Division, CH-1211 Geneva, Switzerland.}\\[3mm]

$^4$~{\it SLAC, Stanford University, Stanford, CA 94309, USA}\\[3mm]
 
$^5$~{\it Fermi National Accelerator Laboratory, P.O.Box 500, Batavia, IL 60510, U.S.A.}\\[3mm]

{\bf Abstract}\\[3mm]
\end{center} 
\setlength{\baselineskip}{0.2in} 
We study logarithmically enhanced electromagnetic corrections to the decay rate in the high dilepton invariant mass region as well as corrections to the forward backward asymmetry (FBA) of the inclusive rare decay $\bar{B} \to X_s
\ell^+ \ell^-$. As expected, the relative effect of these corrections 
in the high dilepton mass region is around $-8$\% for the muonic final state and therefore much larger than in the low dilepton mass region.\\
We also present a complete phenomenological analysis, to improved NNLO accuracy, of the dilepton  mass spectrum and the FBA integrated in the low dilepton mass region, including a new approach to  the zero of the FBA. The latter  represents one of the most precise predictions in flavour physics with a theoretical  uncertainty 
of order 5\%. We find $(q_0^2) _{\mu\mu} = (3.50 \pm 0.12) \gev^2$. For the high dilepton invariant mass region, we have 
${\cal B}(\bar B\to X_s\mu\mu)_{\rm high} = (2.40^{+0.69}_{-0.62} ) \times 10^{-7}$. The dominant  uncertainty is due to the  $1/m_b$ corrections and can be significantly reduced in the future.
For the low dilepton invariant mass region, we confirm previous results
up to small corrections.

\end{titlepage}

\section{Introduction} 
At the beginning of the LHC era, the search for physics beyond the Standard Model (SM) is the main focus of particle physics. In principle, there are two ways to search for possible new degrees of freedom. At the high-energy frontier
one tries to produce those new degrees of freedom directly, while at the high-precision frontier one analyses the indirect virtual effects of such new particles. It is a matter of fact that high-precision measurements allow  to analyse
new physics scales presently not accessible at direct collider experiments.

There are theoretical arguments like the hierarchy problem which let us expect new physics at the electroweak scale. However, the indirect constraints on new physics by the present flavour data indicate a much higher new physics scale when such new effects are naturally parametrized by higher-dimensional operators. Thus, if there is new physics at the electroweak scale, then its flavour structure has to be highly non-trivial and the experimental measurement of flavour-violating couplings is mandatory. This `flavour problem' has to be solved by any new physics scenario at the electroweak scale. Moreover, the present electroweak data also indicate a slightly higher new physics scale (`little hierarchy problem`). 

Among inclusive flavour-changing neutral current (FCNC) processes (for a review see \cite{Hurth:2007xa,Hurth:2003vb}), the inclusive $\bar B \rightarrow X_s \ell^+\ell^-$  decay presents an important test  of the SM, complementary to
the inclusive $\bar B \rightarrow X_s \gamma$ decay.  It is particularly attractive because of  kinematic observables such as the dilepton invariant mass spectrum and the forward-backward asymmetry (FBA). These observables  are 
dominated by perturbative  contributions if the  $c \bar c$ resonances that show up as large peaks in the dilepton invariant mass spectrum are removed by appropriate  kinematic cuts. In the so-called `perturbative $q^2$-windows',
namely in the low-dilepton-mass  region $1\,{\rm GeV}^2 < q^2 = m_{\ell\ell}^2   < 6\,{\rm GeV}^2$, and also in the high-dilepton-mass region with $q^2 > 14.4\,{\rm GeV}^2$, theoretical predictions for the invariant mass spectrum are
dominated by the perturbative contributions, and a theoretical precision of order $10\%$ is in principle possible.

The integrated branching ratio has been measured by both Belle~\cite{Iwasaki:2005sy} and BaBar~\cite{Aubert:2004it} based on a sample of $152 \times 10^6$ and $89 \times 10^6$ $B \bar B$ events respectively. In the low-dilepton invariant mass region, $1\;\gev^2 < q^2 < 6\;\gev^2$, the experimental results read
\bea
{\cal B} (\bar B\to X_s \ell^+\ell^-)_{\rm low} &=& 
\cases{
\left( 1.493 \pm 0.504^{+0.411}_{-0.321} \right) \times 10^{-6} 
& (Belle) \cr 
\left( 1.8 \pm 0.7 \pm 0.5 \right) \times 10^{-6} 
& (BaBar) \cr
\left( 1.60 \pm 0.50 \right) \times 10^{-6} 
& (Average) \cr
}
\eea
Measurements for the high-$q^2$ region, $14.4\;\gev^2 < q^2  < 25\;\gev^2$, are also available:
\bea
{\cal B} (\bar B\to X_s \ell^+\ell^-)_{\rm high} &=& 
\cases{
\left( 0.418 \pm 0.117^{+0.061}_{-0.068} \right) \times 10^{-6} 
& (Belle) \cr 
\left( 0.5 \pm 0.25 ^{+0.08}_{-0.07} \right) \times 10^{-6} 
& (BaBar) \cr
\left( 0.44 \pm 0.12 \right) \times 10^{-6} 
& (Average) \cr
}
\eea
By the end of the present $B$ factories an experimental accuracy of $15\%$ is finally expected.

The recently calculated NNLL QCD contributions~\cite{MISIAKBOBETH,Asa1,Asatryan:2002iy,Adrian2,Asatrian:2002va,Ghinculov:2003bx,Adrian1,Gambinonew,Asatrian:2003yk,Gorbahn:2004my} have significantly improved the sensitivity of the inclusive 
$\bar B \rightarrow X_s \ell^+ \ell^-$ decay in  testing extensions of the SM in the sector of flavour dynamics, in particular, the value of the dilepton invariant mass $q^2_0$ for which the differential FBA vanishes is one of the
most precise predictions in flavour physics with a theoretical uncertainty of order $5\%$. This well corresponds to the expected experimental sensitivity of $4-6 \%$ at the proposed Super-$B$
factories~\cite{Akeroyd:2004mj,Hewett:2004tv,Bona:2007qt}. 

Also non-perturbative corrections scaling with $1/m_b^2$, $1/m_b^3$, or $1/m_c^2$\,\,\cite{Falk,Alineu,Savagenew,Buchalla:1997ky,buchallanewnew,Bauer,Ligeti:2007sn} have to be  taken into account. Moreover, factorizable long-distance contributions away from the resonance
peaks are important; here using the Kr\"uger-Sehgal approach~\cite{KS}  
avoids the  problem of double-counting.

  In the high-$q^2$  region, one encounters the breakdown of the 
heavy-mass expansion at the endpoint; while the partonic contribution vanishes section in the end-point, the $1/m_b^2$ and $1/m_b^3$ corrections tend towards a non-zero value. In contrast to the endpoint region of the photon energy spectrum in the $\bar B \rightarrow X_s \gamma$ decay, no partial all-order resummation into a shape function is possible here. However, for an integrated high-$q^2$ spectrum an effective expansion is found in inverse powers of $m_b^{\rm eff} = m_b \times (1 - \sqrt{\s_{\rm min}})$ rather than $m_b$~\cite{Neubert,BLK}.
The expansion converges less rapidly,  depending on the lower 
dilepton mass cut $\s_{\rm min}$~\cite{Adrian1}. Recently it was suggested~\cite{Ligeti:2007sn} that the large theoretical  uncertainties can be reduced by normalizing the $\bar B \rightarrow X_s \ell^+ \ell^-$ decay rate to the semileptonic  $\bar B \rightarrow X_u \ell\bar\nu$ decay rate with the same $q^2$ cut.

A hadronic invariant-mass cut is imposed in the present experiments (Babar:\, $m_X < 1.8\,{\rm GeV}$, Belle:\, $m_X < 2.0\,{\rm GeV}$) in order to eliminate the background such as   $b \rightarrow  c\, (\rightarrow se^+\nu) e^-\bar \nu = b \rightarrow se^+e^- + \mbox{\rm missing energy}$. The high-dilepton mass region is not affected by this cut, but  in the low-dilepton mass region the kinematics with a jet-like $X_s$ and $m_X^2 \leq m_b \Lambda_{\rm QCD}$ implies the relevance of the 
shape function. 
A recent SCET analysis shows that using the universality of jet and 
shape functions the $10-30\%$ reduction of the dilepton mass spectrum can be accurately computed using the $\bar B \rightarrow X_s \gamma$ shape function. Nevertheless effects of subleading shape functions lead to an additional uncertainty of $5\%$. It is also shown that the zero of the FBA is not affected by the cut~\cite{Lee:2005pk,Lee:2005pw}.
We emphasize that, as in all previous phenomenological analyses of the
inclusive $\bar B \rightarrow X_s \ell^+ \ell^-$ mode, the predictions 
in this paper  are given under the assumption that there 
is no cut in the hadronic mass region.

Finally, as in the $\bar B \rightarrow X_s \gamma$ case, there are unknown subleading nonperturbative corrections
of order $O(\alpha_s \Lambda/m_b)$ which may be estimated by an additional uncertainty of order $5\%$.

Recently, further refinements were presented such as the NLO QED two-loop corrections to the Wilson coefficients whose size is of order $2\%$~\cite{Gambinonew}. Furthermore, it was shown that in the QED one-loop corrections to matrix elements large collinear logarithms of the form $\log(m_b^2/m^2_{\rm lepton})$ survive integration if only a restricted part of the dilepton mass spectrum is considered. This adds another contribution of order $+2\%$ in the low-$q^2$ region for ${\cal B} (\bar B\to X_s \mu^+\mu^-)$~\cite{Huber:2005ig}. For ${\cal B} (\bar B\to X_s e^+e^-)$, in the current BaBar and Belle setups, the logarithm of the lepton mass gets replaced by angular-cut parameters and the integrated branching ratio for the electrons is expected to be close to that for the muons. 

In the present manuscript, we calculate  these  corrections due to large collinear logarithms  also for the dilepton mass spectrum in the high-$q^2$ window and for the FBA for the first time. In the high-$q^2$ region these corrections
are much larger than in the low-$q^2$ one; we already anticipate that in the muon channel the former account for $-8\%$ of the rate while the latter are only about $+2\%$. We also update the theoretical prediction of the observables using several
improvements offered recently in the literature among which the two-loop matrix element of the  semileptonic operator $P_9$ and the $1/m_b^3$ corrections are the most important to date. We present a new phenomenological analysis of the
dilepton mass spectrum in the low- and  high-$q^2$ region and of the FBA. In particular, a new approach to the zero of the FBA is proposed. 
Moreover, we explore a strategy to reduce the uncertainties due to the $1/m_b$ corrections by 
normalizing with the semileptonic  $\bar B \rightarrow X_u \ell\bar\nu$ decay rate with the same $q^2$ cut, as was recently proposed in the literature~\cite{Ligeti:2007sn}.
We already anticipate that with this normalisation the dilepton mass spectrum,  integrated over  the high-$q^2$  region, receives only a $3\%$ shift in the central value by the nonperturbative corrections.

Finally, we derive  model-independent formulae for non-SM values of the high-scale Wilson coefficients.   

The present manuscript is organised as follows. In Section~\ref{sec:MF} we present the master formulae for the calculation of the observables; in particular we discuss the normalization of the branching ratio and forward backward
asymmetry and explain the details of the perturbative expansion. In Section~\ref{sec:analyticalresults} we present the novel analytical results on the logarithmically enhanced QED corrections to the forward backward asymmetry and to the
decay width in the high-$q^2$ region and also discuss some nonperturbative subtleties. Furthermore, we analyse  the relation of the collinear logarithms 
and the  experimental angular cuts. In  Section~\ref{sec:results} we present the SM predictions 
for the branching ratio and forward--backward asymmetry.  In Section~\ref{sec:conclusions} we draw our conclusions.

\section{Master formulae}
\label{sec:MF}
In this section, we present master formulae for the various observables. We closely follow here the notations of Ref.~\cite{Huber:2005ig} in order to keep the paper compact.
The new results are explicitly derived in the next section. 

\subsection{Dilepton invariant mass spectrum}
\label{sec:masterDMS}
Following Ref.~\cite{Huber:2005ig}, the master formula for the $\bar B \to X_s \ell^+\ell^-$ branching ratio reads
\bea
{{\rm d} {\cal B} (\bar B\to X_s \ell^+\ell^-) \over {\rm d} \hat s} & = &
{\cal B} (\bar B\to X_c e \bar\nu)_{\rm exp} \; 
\left| \frac{V_{ts}^* V_{tb}}{V_{cb}} \right|^2 \; 
\frac{4}{C} \; \frac{\Phi_{\ell\ell}(\s)}{\Phi_u} \;, 
\label{br}
\eea
where $\s = q^2/m_{b,{\rm pole}}$. $\Phi_{\ell\ell}(\s)$ and $\Phi_u$ are
defined by
\bea \label{bu}
\frac{{\rm d}\Gamma(\bar B \to X_s \ell^+\ell^-)}{{\rm d}\s} &=& \frac{G_F^2 m_{b,{\rm pole}}^5}{48\pi^3} \left| V_{ts}^{\ast}V_{tb}^{}\right|^2 \; \Phi_{\ell\ell}(\s) \; ,\\
\Gamma (\bar B\to X_u e\bar\nu) &=&
\frac{G_F^2 m_{b,{\rm pole}}^5}{192 \pi^3} \left| V_{ub}^{}\right|^2 \; \Phi_u \; ,
\eea
and we have $\Phi_u = 1 + {\cal O}(\alpha_s,\alpha_{\mathrm em},\Lambda^2/m_b^2)$. The normalization with the measured semileptonic decay rate minimizes 
the uncertainty due to the fifth power of $m_b$; the factor C,  
\be
C = \left| \frac{V_{ub}}{V_{cb}} \right|^2 
       \frac{\Gamma (\bar B\to X_c e\bar\nu)}{\Gamma (\bar B\to X_u e\bar\nu)} \;,
\ee
is used to avoid spurious uncertainties due to the $b \rightarrow X_c e\bar \nu$ phase-space factor. This factor can be determined from a global analysis of the semileptonic data~\cite{Bauer:2004ve}. The quantity $\Phi_{\ell\ell}(\s)/\Phi_u$ can be expressed in terms of the low-scale Wilson coefficients and various functions of $\s$ that arise from the matrix elements. The main formula reads
\bea \label{mainmaster}
\f{\Phi_{\ell\ell}(\s)}{\Phi_u} &=& \sum_{i\leq j} 
{\rm Re} \left[ C_i^{\rm eff} (\mu_b) \; C_j^{{\rm eff}*} (\mu_b)  
                \; H_{ij} (\mu_b,\s) \right] \;,
\eea
where $C_i^{\rm eff}(\mu_b) \neq C_i (\mu_b)$ only for $i=7,8$. The functions $H_{ij}(\mu_b,\s)$ are given in Eq.~(116) of Ref.~\cite{Huber:2005ig} and can be expressed in terms of the coefficients $M_i^A$ listed in Table 6 of Ref.\cite{Huber:2005ig}, as well as the building blocks $S_{99}$, $S_{77}$, $S_{79}$, and $S_{1010}$ given in Eqs.~(112--115) of Ref.\cite{Huber:2005ig}.

In the low-$q^2$ region, there are only a few modifications compared to the analysis in Ref.~\cite{Huber:2005ig}: we update the parametric inputs, the influence of the $c \bar c$ resonances are taken into account and the two-loop matrix element of the operator $P_9$ is given in a more precise way (see Eqs.~(\ref{omega99}) and~(\ref{cbarc}) of the present paper). The combined effect of these improvements does not change appreciably the central value for the branching ratio obtained in Ref.~\cite{Huber:2005ig}. 

In the high-$q^2$ region, there are several additional changes necessary besides the ones mentioned above. The electromagnetic contributions to the building blocks $S_{99}, S_{77},S_{79}$, and $S_{1010}$ are different and are presented
here for the first time in Eqs.~(\ref{omega29}) --~(\ref{sigma6}). $O(1/m_b^3)$ corrections are numerically relevant and explicitly given in  Eqs.~(6) of Ref.~\cite{Ligeti:2007sn}. Finally, the functions
$F_1^9$, $F_2^9$, $F_1^7$, $F_2^7$, $F_8^9$, and $F_8^7$ that appear in the coefficients  $M_i^A$ (listed in Table 6 of Ref.\cite{Huber:2005ig}) are different in the high-$q^2$ region and are known only numerically. They have been
calculated in Ref.~\cite{Adrian1}.\footnote{Numerical expressions 
of these functions can be obtained from the authors upon request.}

\subsubsection{New normalization}\label{sec:R0}
The authors of Ref.\cite{Ligeti:2007sn} have shown that it is possible to drastically reduce the size of $1/m_b^{2}$ and $1/m_b^{3}$ power corrections to the integrated decay width, by normalizing it to the semileptonic $b\to u \, \ell
\, \nu$ rate integrated over the same $q^2$-interval. This procedure will help reducing the uncertainties induced by the large power corrections to the decay width integrated over the high-$s$ region. The new observable is defined as
follows:
\bea\label{eq:zoltanR}
{\cal R}(s_0) & = & \frac{\displaystyle
\int_{\hat s_0}^1 {\rm d} \hat s \, {{\rm d} {\Gamma} (\bar B\to X_s \ell^+\ell^-) \over {\rm d} \hat s}
}{\displaystyle
\int_{\hat s_0}^1 {\rm d} \hat s \, {{\rm d} {\Gamma} (\bar B^0\to X_u \ell \nu) \over {\rm d} \hat s}
} 
=
4 \left| V_{ts} V_{tb} \over V_{ub} \right|^2 \frac{
\int_{\hat s_0}^1 {\rm d} \hat s \, \Phi_{\ell\ell} (\hat s)}{
\int_{\hat s_0}^1 {\rm d} \hat s \, \Phi_{u} (\hat s)
} 
\eea
where $\Phi_{\ell\ell}(\s)$ was defined in Eqs.~(\ref{br}) and~(\ref{bu}) and the differential $\Phi_u (\hat s)$ is given by
\bea
\frac{{\rm d} \Gamma_u}{{\rm d} \hat s} = 
\frac{G_F^2 \left| V_{ub} \right|^2 m_{b,{\rm pole}}^3}{192 \pi^3} \; \Phi_u (\hat s) \; .
\eea
Explicit expressions for the $O(1,\alpha_s,1/m_b^2,1/m_b^3)$ contributions to $\Phi_u(\hat s)$ can be found in Eqs.~(5,11) of Ref.~\cite{Ligeti:2007sn}. The $O(\alpha_s^2)$ correction to $\Phi_u(\hat s)$
is given by $\alpha_s^2 X_2 (\hat s)$ with $X_2$ defined in Eq.~(60) of Ref.~\cite{semi3}. We have not included electromagnetic corrections of order $O(\kappa)$ in the normalization because they are presently not known. However, the impact
of those effects in the conventional approach where the fully integrated semileptonic decay rate is used as normalization is less than $2.5$\%.
We would also like to note that in Eq.~(\ref{eq:zoltanR}), 
contrary to Eq.~(\ref{br}), corrections of ${O}(\alpha_s \kappa)$ 
to $\Phi_u(\hat s)$ can contain residual logarithmically enhanced terms once the integration over
$\hat s$ is restricted to the high-$q^2$-region. We leave the investigation of terms of $O(\kappa)$ and ${O}(\alpha_s \kappa)$ to a future analysis.

Note that in the definition of $\cal R$ we used only neutral 
semileptonic decays in order to reduce the 
uncertainties coming from weak annihilation contributions 
(see further remarks in Section 4.1).

Obviously, for the comparison of the theoretical ratio $\cal R$ with experiment, a separate measurement of $B^0 \rightarrow X_u \ell\bar\nu$ and $B^{\pm} \rightarrow X_u\ell\bar\nu $ 
with a high-$q^2$ cut is necessary. These are quantities with larger experimental uncertainties. However, they 
are not expected to negate the drastic reduction of the theoretical uncertainty in the ratio $\cal R$ (see section~\ref{sec:R0res} for numerical results).

\subsection{Forward-backward asymmetry}
\label{sec:masterFBA}
In complete analogy to the formula for the branching ratio, one can derive also a formula
that expresses the FBA (see \cite{Ali:1991is}) in terms of the 
low-scale Wilson coefficients and various building blocks. Normalizing to the semileptonic $B \to X_u e \bar\nu$ decay width, we get
\bea
{{\rm d} {\cal A}_{FB} (\bar B\to X_s \ell^+\ell^-) \over {\rm d} \hat s} & = &
{\cal B} (B\to X_c e \bar\nu)_{\rm exp} \; 
\left| \frac{V_{ts}^* V_{tb}}{V_{cb}} \right|^2 \; 
\frac{4}{C} \; \frac{\Phi^{FBA}_{\ell\ell}(\s)}{\Phi_u} \;, 
\label{fba}
\eea\
where $\Phi^{FBA}_{\ell\ell}(\s)$ is defined by ($z=\cos\theta_\ell$)
\bea\label{eq:doublediff}
\int\limits_{-1}^1\! {\rm d}z \, {\rm sgn}(z) \, \frac{{\rm d}^2 \Gamma(\bar B \to X_s \ell^+\ell^-)}{{\rm d} \hat s \, {\rm d}z}&=&\frac{G_F^2 m_{b,{\rm pole}}^5}{48\pi^3} \left| V_{ts}^{\ast}V_{tb}^{}\right|^2 \;
\Phi^{FBA}_{\ell\ell}(\s) \; ,
\eea
%We write as in Ref.~\cite{Huber:2005ig}:  
\bea \label{mainmasterAFB}
\dps \frac{\Phi^{FBA}_{\ell\ell}}{\Phi_u} &=& \sum_{i\leq j} 
\; {\rm Re} \left[ C_i^{\rm eff} (\mu_b) \; C_j^{{\rm eff}*} (\mu_b)  
                \; H_{ij} (\mu_b,\s) \right] \;.
\eea
Again, the functions $H_{ij}(\mu_b,\s)$ depend on the coefficients $M_i^A$ listed in Table~6 of Ref.~\cite{Huber:2005ig} and on the building blocks $S_{710}$ and $S_{910}$: 
\be 
H_{ij} = \left\{ \begin{array}{ll}
{}~\sum~~ \;{\rm Re} (M_i^A M_i^{10*}) \;S_{A10} +\Delta H_{ii}\;,
& \mbox{when~} i=j \\[-2mm]
{\,\scriptscriptstyle A=7,9}\\[2mm]
{}~\sum~~ \; \left(M_i^A M_j^{10*} + M_i^{10} M_j^{A*} \right) \; S_{A10} +\Delta H_{ij}\;,
& \mbox{when~} i\neq j \; .\\[-2mm]
{\,\scriptscriptstyle A=7,9}
\end{array}\right.\label{hijAFB}
\ee 
where $S_{710}$ and $S_{910}$ are given by 
\bea
{\rm S}_{710} & = & -6\,(1-\s)^2 \left\{ 
                   1 + 8\; \as \, \left[ f_{710} (\s) +u^{(1)}\right] + \kappa \, u^{(\rm em)}
                     + 8\; \as \kappa \; \left[ \omega_{710}^{(\rm em)} (\s)+ u^{(\rm em)} \,  f_{710} (\s)\right]\right. \nnb \\ &&
		\hspace*{65pt}     \left. + 16 \, \as^2 \left[u^{(2)} + 4 \; u^{(1)} \, f_{710} (\s)\right] \right\}\nnb \\ &&
-\frac{8 \, \lambda_1}{m_b^2}\; \s -\; 6 \; \frac{\lambda_2}{m_b^2} \; (1-14 \s+ 9\s^2) \;,\\
{\rm S}_{910} & = & -3 \, \s \,  (1-\s)^2 \left\{1+ 8\; \as \, \left[ f_{910} (\s) +u^{(1)}\right] + \kappa \, u^{(\rm em)} 
                   + 8\; \as \kappa \; \left[ \omega_{910}^{(\rm em)} (\s) + u^{(\rm em)} \,  f_{910} (\s)\right]\right. \nnb \\ &&
		\hspace*{70pt}    \left. + 16 \, \as^2 \left[u^{(2)} + 4 \; u^{(1)} \, f_{910} (\s)\right] \right\}\nnb \\ &&
-\frac{4 \, \lambda_1}{m_b^2}\; \s^2+\, \frac{12\,\lambda_2}{m_b^2} \, \s^2 \, (4-3\s) \;.\label{eq:s710BR} 
\eea
The functions $f_{i10}(\s)$ ($i=7,9$) can be found in Eqs.~(15--17) of Ref.~\cite{Asatrian:2002va}. The functions $\omega_{ij}^{\rm (em)}$ represent the electromagnetic corrections to the matrix elements which are calculated for the first time in the present paper (see Eqs.~(\ref{omega710em}--\ref{sigma7})). The $S_{AB}$ also include non-perturbative ${\cal O}(1/m_b^2)$ corrections from Refs~\cite{Alineu,buchallanewnew}.  Contrary to the expression for the branching ratio the quantity $\lambda_1$, which is related to the kinetic energy of the $b$-quark, does not drop out here. The quantities
\be
\Delta H_{ij} = b_{ij} + c_{ij} + e_{ij} 
\ee
have the same meaning as in Eq.~(117) of Ref.~\cite{Huber:2005ig}. They need to be included only for $i=1,2$. The additional $\ln (m_b^2/m_\ell^2)$-enhanced electromagnetic corrections $e_{ij}$ for the FBA read
\bea
e_{210} & = & -24\,\s\,(1-\s)^2\;\as^2 \kappa^2\; \omega_{210}^{\rm (em)}(\s) \nnb\\
e_{110} & = &  \fm{4}{3}  e_{210}.
\eea
The function  $\omega_{210}^{\rm (em)}(\s)$ is again new (see Eq.~(\ref{omega210em})), while the ${\cal O}(\Lambda_{\rm QCD}^2/m_c^2)$ non-perturbative contributions were calculated in
Ref.~\cite{Buchalla:1997ky}
\bea
c_{210} &=& +\as \kappa  \frac{\lambda_2}{3 m_c^2} (1-\s)^2
                  (1+3\s)\,%  {\rm Re}\,
		  F(r)  \;,\nnb\\
c_{110} & = &  -\fm{1}{6}  \; c_{210} \; , 
\eea 
where $r \equiv 1/y_c = s/(4 m_c^2)$ and the function  $F(r)$  is listed for example in appendix~A of Ref.~\cite{Huber:2005ig}. The finite bremsstrahlung contributions $b_{ij}$ were calculated in Ref.~\cite{Asatrian:2003yk}. We do not present these corrections here but do include them in the numerical analysis. 

\subsection{Normalization of the FBA}
As discussed in section 2, the normalization of the FBA by the semileptonic or by the $\bar B\rightarrow X_s \ell^+ \ell^-$ decay rate is important in order to cancel the overall factor $m_{b, {\rm pole}}^5$. Moreover, the elimination of renormalon ambiguities requires the analytical conversion to a short-distance mass and a complete expansion in $\alpha_s$~\cite{Hoang:1998hm,Hoang:2000fm}. 

If we extract the zero of the FBA without using any normalization, the conversion and the  expansion of the overall $m_{b, {\rm pole}}^5$ factor induces a large dependence on the scheme we choose for the short distance $b$ mass. In fact, we find a $-11\%(+5\%)$ shift in switching from the 1S scheme to the $\overline{\rm MS}$ (pole) one. This large scheme $m_b$ dependence was overlooked in previous analyses  of the zero. It  can be eliminated by normalizing the FBA by a rate proportional to $m_b^5$ and fully expanding in $\alpha_s$.

The most natural choice is to normalize the FBA directly by the differential $\bar B \rightarrow X_s \ell^+ \ell^-$ differential decay rate  
because this normalisation is also used in the experiment.
Unfortunately,  this option is not viable because of the poor convergence of 
the perturbative series for the latter (the most serious problem is that the lowest term in the expansion
is sub--dominant). It is therefore not surprising that also this procedure leads to a large $m_b$ scheme dependence of the zero: we find a shift of $-12\%$ when one compares the 1S to the $\overline{\rm MS}$ scheme. 

Therefore,  we propose here a definition of the FBA that, at the same time, removes the $m_b^5$ dependence, does not suffer of renormalon ambiguities, has a nicely behaved perturbative expansion, and can directly compared with the experimental quantity.  The following double ratio satisfies all the requirements:
\be \label{doubleratio}
[\frac{{\rm d} {\cal A}_{FB} (\bar B\to X_s \ell^+\ell^-)}{{\rm d} \hat s}]\, /\, [\frac{{\rm d} {\cal B} (\bar B\to X_s \ell^+\ell^-)}{{\rm d} \hat s}]\, =\,   [\frac{\Phi^{FBA}_{\ell\ell}(\s)}{\Phi_u}]\,/\,[\frac{\Phi_{\ell\ell}(\s)}{\Phi_u}]\,,  
\ee
where the quantities are defined in Eqs. (\ref{br}) and (\ref{fba}). The critical point is that each of the ratios in brackets gets fully expanded in $\alpha_s$, but no overall expansion is done. This is allowed for two reasons: on the one side, the fully expanded numerator and denominator are renormalon free, on the other one, they separately correspond to two measurable quantities. As we show in Sec.~\ref{sec:analysezeroFBA}, also the $m_b$ scheme dependence of the zero turns out to be minimal (the difference between the 1S, $\overline{\rm MS}$ and pole schemes is about $1\%$).

This discussion shows that the estimation of the perturbative error of the zero via the comparison of different normalisations as done in~\cite{Gambinonew} is questionable because some normalizations are affected by renormalon ambiguities and the problem of the large scheme dependence of the $b$ quark mass should be addressed. As we will discuss in section~\ref{sec:analysezeroFBA}, the standard analysis of the scale dependence offers a better guidance to the understanding of the perturbative errors in our predictions.

\subsection{Perturbative expansion} 
\label{perturbative}
Some remarks about the perturbative expansion are in order: Large logarithms of the form $\alpha_s \log(m_b/M_W)$ have to be resummed at a given order in perturbation theory using renormalization group techniques. In the case of $b \rightarrow s \ell^+\ell^-$, the first large logarithm of the form $\log(m_b/M_W)$ arises already without  gluons. Moreover, the amplitude of $b \rightarrow s \ell^+ \ell^-$ is proportional to $\alpha_{\rm em}$. So naturally we have an expansion in $\alpha_s$ and in $\kappa=\alpha_{\rm em}/\alpha_s$  while resumming all powers of $\alpha_s \log(m_b/M_W)$. The Leading Order (LO) are of order $\kappa$, NLO are proportional $\kappa\alpha_s$, NNLO are proportional $\kappa \alpha_s^2$.

It is well-known that this naive $\alpha_s$ expansion is problematic, since the formally-leading $O(1/\alpha_s)$ term in $C_9$ is accidentally small and much closer in size to an $O(1)$ term. In addition the NLO terms are enhanced by a factor of $m_t^2/(M_W^2 \sin^2\theta_W)$. Therefore, also specific higher order terms in the general $\kappa^n \alpha_s^m$ expansion are numerically important.

The $b\to s \ell^+\ell^-$ decay amplitude has the following structure (up to an overall factor of $G_F$):
\bea
{\cal A} & = & \hspace{5mm} \kappa \left[ {\cal A}_{LO}  + \alpha_s  \; {\cal A}_{NLO}+  
               \alpha_s^2 \; {\cal A}_{NNLO} + {\cal O}(\alpha_s^3) \right]
               \nonumber\\
         &   & +\kappa^2 \left[
               {\cal A}_{LO}^{\mathrm em} + \alpha_s \; {\cal A}_{NLO}^{\mathrm em}  + 
               \alpha_s^2  \; {\cal A}_{NNLO}^{\mathrm em} + {\cal O}(\alpha_s^3) \right] 
\; + {\cal O}(\kappa^3) \;.
\eea
with ${\cal A}_{LO} \sim  \alpha_s  \; {\cal  A}_{NLO}$ and
${\cal A}_{LO}^{\mathrm em} \sim \alpha_s \; {\cal A}_{NLO}^{\mathrm em}$.
All these terms are included in the numerical  analysis in a complete manner, together
with the appropriate bremsstrahlung corrections, while also the term ${\cal A}_{NNLO}$ is practically complete due to the calculations in Refs.~\cite{MISIAKBOBETH,Asa1,Asatryan:2002iy,Adrian2,Asatrian:2002va,Ghinculov:2003bx,Adrian1,Gambinonew,Asatrian:2003yk,Gorbahn:2004my}. The only missing parts originate from the unknown two-loop matrix elements of the QCD-penguin operators whose Wilson coefficients are very small. 

Among the contributions to ${\cal A}_{NNLO}^{\mathrm em}$, we include only the terms which are either enhanced by an additional factor of $m_t^2/(M_W^2\sin^2\theta_W)$ (with respect to ${\cal A}_{NLO}^{\mathrm em}$) \cite{Gambinonew} or contribute to the $\ln (m_b^2/m_\ell^2)$-enhanced terms at the decay width level. As discussed above, the latter terms were  calculated in Ref.~\cite{Huber:2005ig} for the dilepton mass spectrum in the low-$q^2$ region, while the corresponding terms for the high-$q^2$ region and the FBA are calculated for the first time in the present paper.

The perturbative expansion of the ratio $\Phi_{\ell\ell}(\s)/\Phi_u$ has  a similar  structure  to  that of the squared amplitude (up to bremsstrahlung corrections and nonperturbative corrections)
\bea
{\cal A}^2 
& = & 
\kappa^2 \Big[ {\cal A}_{LO}^2 
+  \alpha_s \; 2 {\cal A}_{LO} {\cal A}_{NLO} 
+ \alpha_s^2 \; ({\cal A}_{NLO}^2 + 2  {\cal A}_{LO} {\cal A}_{NNLO}  ) 
\nonumber\\
& & \hskip 0.5 cm 
+  \alpha_s^3 \; 2 ({\cal A}_{NLO} {\cal A}_{NNLO} + \ldots)  + {\cal O}(\alpha_s^4))
\Big]   \nonumber \\
&+ &
\kappa^3 \Big[
2 {\cal A}_{LO}^{} {\cal A}_{LO}^{\mathrm em} + 
\alpha_s \; 2 ({\cal A}_{NLO}^{} {\cal A}_{LO}^{\mathrm em} + {\cal A}_{LO}^{} {\cal A}_{NLO}^{\mathrm em}) 
\nonumber\\
& &
\hskip 0.5 cm 
+\alpha_s^2 \; 2 ( {\cal A}_{NLO}^{} {\cal A}_{NLO}^{\mathrm em}
+ {\cal A}_{NNLO}^{} {\cal A}_{LO}^{\mathrm em} 
+ {\cal A}_{LO}^{} {\cal A}_{NNLO}^{\mathrm em}) \nonumber\\
&& \hskip 0.5 cm
+ \alpha_s^3 \; 2 ({\cal A}_{NLO}^{} {\cal A}_{NNLO}^{\mathrm em} + {\cal A}_{NNLO}^{} {\cal A}_{NLO}^{\mathrm em}
+ \ldots)
+ {\cal O}(\alpha_s^4)\Big] \nonumber \\
&+ &
 {\cal O}( \kappa^4)\; .
\label{schematic2}
\eea
We assume that all products in Eq.~(\ref{mainmaster}) are expanded in  $\alpha_s$ and $\kappa$. Regarding QCD, a strict NNLO calculation
of $\Phi_{\ell\ell}(\s)/\Phi_u$ should only include terms up to order $\kappa^2 \alpha_s^2$.

In the  numerical calculation of
$\Phi_{\ell\ell}(\s)/\Phi_u$, however, we include all the terms that are written explicitly in  Eq.~(\ref{schematic2}). The term ${\cal
A}_{NLO} {\cal A}_{NNLO}$ of  order $\kappa^2 \alpha_s^3$ is  formally a NNNLO term, but numerically important. Within the
electromagnetic corrections the same is true for the term ${\cal A}_{NLO}^{\mathrm em} {\cal A}_{NNLO}$. We emphasize that those terms beyond the formal NNLO level which are proportional to $|C_7|^2$ and $|C_8|^2$ are scheme independent. 
The dots in   Eq.~(\ref{schematic2}) stand for the unknown terms ${\cal A}_{LO} {\cal A}_{NNNLO}$ and ${\cal A}_{LO}^{\mathrm em} {\cal A}_{NNNLO}$ and, consequently, can safely be neglected due to ${\cal A}_{LO} \sim  \alpha_s  {\cal  A}_{NLO}$, ${\cal A}_{LO}^{\mathrm em} \sim \alpha_s {\cal A}_{NLO}^{\mathrm em}$, and $\alpha_s {\cal A}_{NNNLO} \ll  {\cal A}_{NNLO}$. Thus, 
one can argue that our calculations pick up the dominant NNNLO QCD corrections.
In the following we will therefore call the accuracy of our calculations {\it improved} NNLO.~\footnote{In  fact, this scheme  is very similar to the
one  proposed in Ref.~\cite{Asa1} and used in many previous analyses of the $b \rightarrow s \ell^+ \ell^-$ observables. In these
works the formally-leading, but accidentally small $O(1/\alpha_s)$ term in $C_9$ is treated as $O(1)$ and absorbed into the NLO
coefficient: as consequence the two-loop matrix element of $P_9$ and the three-loop mixing of the four-quark operators into $P_9$ are
left out even though they are formally NNLO terms. This is the main difference between the NNLO scheme proposed in Ref.~\cite{Asa1} and
the {\it improved} NNLO used in the present manuscript.}

Finally let us comment on the mass schemes that we use:  The pole mass of the $b$ quark appears explicitly in the calculation of the
matrix elements ($\hat s \equiv s/m^2_{b, {\rm pole}}$) and in several loop functions. Unfortunately, the precise determination of the
numerical value of $m^2_{b, {\rm pole}}$ is hindered by renormalon ambiguities that appear in the relation between any short distance
mass definition (e.g. $\overline{\rm MS}$, 1S, kinetic schemes) and the pole one. We eliminate these renormalon uncertainties utilizing the Upsilon
expansion described in Refs.~\cite{Hoang:1998hm,Hoang:2000fm}. Every occurrence of the pole mass is converted analytically to the
$1S$-mass before any numerical evaluation of the branching ratio is performed. In our analysis we use the conversion formula up to order
$\alpha_s^2$~\cite{Hoang:2000fm}. We follow a similar approach for the treatment of the charm pole mass that appears in the calculation
of some matrix elements. We adopt the $\overline{\rm MS}$ charm mass as input and expand the pole mass at order $\alpha_s^2$ using the formulae
presented in Ref.~\cite{Hoang:2000fm}. For what concerns the top mass, we take the pole mass as input and convert it to the $\overline{\rm MS}$ scheme
at order $\alpha_s^3$. The perturbative expansion of the FBA proceeds along the same lines. 

\section{Analytical  results}
\label{sec:analyticalresults}
In this section, we calculate the electromagnetic corrections to the matrix elements for the FBA in the low- and high-$q^2$ region and
for the dilepton mass spectrum in the high-$q^2$ region 
and discuss the phenomenological implications of those collinear 
logarithms in the $ee$ and $\mu\mu$ final states.

Moreover, we present a more precise fixing  of the two-loop matrix elements of
the $P_9$ operator. Finally, we make some remarks on some non--perturbative subtleties. 

\subsection{Log-enhanced corrections to the FBA, low- and high-$q^2$ region}
\label{sec:FBAlowsandhighs}
First we derive the basic formulae to obtain the expressions for the $\ln(m_b^2/m_{\ell}^2)$-enhanced corrections to the FBA, where we
shall focus here on the un-normalized asymmetry\footnote{In this subsection 
we  suppress the normalization factor, thus,
the symbol ${\cal A}_{FB}$ denotes the un-normalized FBA, see
Eqs.~(\ref{eq:fullydiffFBA}),~(\ref{eq:Asquarenotation}), and~(\ref{deltaFBA}).} corresponding to the LHS of Eq.~(\ref{eq:doublediff}). We follow closely the notation of Section 5 of Ref.~\cite{Huber:2005ig}.

It was shown in the Appendix of Ref.~\cite{Alineu} that the angular FBA with respect to $\theta_{\ell}$, the angle in the dilepton
c.m.s. between the directions of the momenta of the decaying $\bar B$ and the positively charged lepton, is equivalent to the energy
asymmetry between the two leptons in the rest--frame of the decaying $\bar B$. Events in which $\cos\theta_{\ell}>0$ in the dilepton
c.m.s. correspond to events in which $E_{-} > E_{+}$ measured in the $\bar B$-meson restframe, where $E_{\pm}$ denotes the
$\ell^{\pm}$-energy. Sticking to the latter frame and
defining the scaled energies
\be
\dps y_{\pm} \equiv \frac{2 \, E_{\pm}}{m_b} \; ,
\ee
we can write the fully differential FBA as
\be\label{eq:fullydiffFBA}
d{\cal A}_{FB} = PF \; d\s \; dy_{+} \; dy_{-} \; \delta(1+\s-y_{+}-y_{-}) \, \left|{\cal A}\right|^2 \, {\rm sgn}(y_{-}-y_{+}),
\ee
with the pre-factor 
\be
\dps PF = \frac{G_F^2 m_b |V_{tb} V_{ts}^{\ast}|^2}{32 \pi^3} \; .
\ee
We are mainly interested in $\ln(m_b^2/m_{\ell}^2)$-enhanced electromagnetic corrections. They are derived by means of the splitting function and we shall adopt the kinematics from Figure 2 of Ref.~\cite{Huber:2005ig}. In the collinear
limit the fully differential FBA reads
\be
d{\cal A}^{(m)}_{FB,{\rm coll}} =  PF  \; dx \; d\s \; dy_{+} \; dy_{-} \; \delta(1+\s-y_{+}-y_{-}) \, f^{(m)}_{\gamma}(x) \, \left|{\cal
A}\right|^2 \, {\rm sgn}(y_{-}-y_{+}) \;. \label{AFBfully}
\ee
%
%with the pre-factor 
%
%\be
%\dps PF = \frac{G_F^2 m_b |V_{tb} V_{ts}^{\ast}|^2}{32 \pi^3} \; .
%\ee
%
We shall only retain the $\ln(m_b^2/m_{\ell}^2)$-enhanced part of $\dps f^{(m)}_{\gamma}$, which then becomes independent of $E$,
\be
f^{(m)}_{\gamma}(x) = 4 \, \tilde{\alpha}_e \, \frac{[ 1+(1-x)^2]}{x} \, \ln\!\lk\frac{m_b}{m_{\ell}}\rk \; .
\ee
The approximation of dropping the non-log-enhanced terms in the splitting function is justified by the fact that even for high $q^2$ the logarithmically enhanced part dominates numerically. This can be seen, for instance, by
the fact that the non-log-enhanced term of $(1-\hat s)^2 \, (1+2\hat s) \, \omega^{(\rm em)}_{99}(\hat s)$ (see Eq.~(94) of Ref.~\cite{Huber:2005ig}) vanishes at $\hat s=1$.

In the squared amplitude in Eq.~(\ref{AFBfully}) we keep only those terms that are relevant for the FBA, i.~e.\ all terms that do not drop out upon integration over the sign-function,
\bea
\dps \left|{\cal A}\right|^2 &=& 2 \; {\rm Re} \Big[ C_{10} C_9^* + \as \kappa \; C_{10} (C_2 + C_F \; C_1) f_2^* (\s)\Big]
\; \me{P_{10}}_{\rm tree}\me{P_9}_{\rm tree}^*  \nnb\\
&&+ 2 \; {\rm Re} [ C_{10} C_7^*] \; \me{P_{10}}_{\rm tree}\me{P_7}_{\rm tree}^* \; .\label{eq:Asquarenotation}
\eea
As in the case of the decay width we must consider the difference
\be\label{doubleminustripleAFB}
\dps \frac{d{\cal A}^{(m)}_{FB,{\rm coll},2}}{d\s} - \frac{d{\cal A}^{(m)}_{FB,{\rm coll},3}}{d\s} \; ,
\ee
where we stay differential in the double and in the triple invariant, respectively. At this point more care is required compared to the calculation of the decay width. Due to the emergence of the sign-function in Eq.~(\ref{AFBfully}) we must distinguish between photon-emission from the $\ell^+$ and from the $\ell^-$ in case of the double invariant. In the former case we have to stay differential in $\s=(\bar x p_1+p_2)^2/m_b^2$, in the latter in $\s=(p_1+\bar x p_2)^2/m_b^2$. Also the $y_{\pm}$ change accordingly. We therefore have
\bea\label{doubleplusminus}
\dps \frac{d{\cal A}^{(m)}_{FB,{\rm coll},2,\ell^{\pm}}}{d\s} &=& \pm PF \lek \int\limits_0^{1-\sqrt{\s}}\!\!\!dx \int\limits_{\s/\bar
x}^{\frac{\bar x+\s}{\bar x(1+\bar x)}}\!\!\!dy_{\pm} + \int\limits_{1-\sqrt{\s}}^{1-\s}\!\!\!dx \int\limits_{\s/\bar
x}^{1}\!\!\!dy_{\pm}- \int\limits_0^{1-\sqrt{\s}}\!\!\!dx \int\limits_{\frac{\bar x+\s}{\bar x(1+\bar x)}}^{1}\!\!\!dy_{\pm} \rek \nnb\\
&&\nnb\\
&&\times \, \frac{f^{(m)}_{\gamma}(x)}{\bar x} \, \left|{\cal A}\right|^2 {}_{\big| \s \to \s/\bar x \; ; \; y_{\mp} \to 1 - y_{\pm} +
\s/\bar x} \quad .
\eea
The two expressions corresponding to upper and lower sign should be equal due to the antisymmetry of $\left|{\cal A}\right|^2$ in $y_{+}
\leftrightarrow y_{-}$. The case of the triple invariant is simpler since we stay differential in  $\s=(p_1+p_2)^2/m_b^2$, namely
\be\label{tripleplusminus}
\dps \frac{d{\cal A}^{(m)}_{FB,{\rm coll},3,\ell^{\pm}}}{d\s} = \pm PF \int\limits_0^1\!dx \lek
\int\limits_{\s}^{\frac{1+\s}{2}}\!\!\!dy_{\pm} - \int\limits_{\frac{1+\s}{2}}^1\!\!\!dy_{\pm}\rek
\, f^{(m)}_{\gamma}(x) \, \left|{\cal A}\right|^2 {}_{\big| \; y_{\mp} \to 1 - y_{\pm} +\s} \quad ,
\ee
where the expression for the upper and lower sign should again be equal due to the antisymmetry of $\left|{\cal A}\right|^2$. We finally have to combine the expressions according to Eq.~(\ref{doubleminustripleAFB}).

The corrections to the unnormalized FBA read
\bea
\frac{{\rm d} \Delta {\cal A}_{FB}}{d\s} &=& \frac{G_F^2 m_{b}^5}{48\pi^3} \left| V_{ts}^{\ast}V_{tb}^{}\right|^2 \; \Delta
\Phi^{FBA}_{\ell\ell}(\s) \nnb \\
&=&\frac{G_F^2 m_b^5}{48 \pi^3} |V_{tb} V_{ts}^{\ast}|^2 (1-\s)^2 \as \kappa %\ln \frac{m_b^2}{m_\ell^2} 
\Bigg\{ - 48 \Bigg[ \as \kappa \, 
{\rm Re} \left[C_7 C_{10}^*\right] \;  \omega_{710}^{\rm (em)}(\s) \Bigg]\nnb\\
%& &
%+\as\kappa \; {\rm Re} \left[(C_2 + C_F C_1) C_9^* \;  \omega_{29}^{\rm (em)}(\s) \right]
%+\as^2 \kappa^2 \; (C_2 + C_F C_1)^2 \;  \omega_{22}^{\rm (em)} (\s)
%\Bigg]
%\nnb \\
& & 
- 24 \, \s \, \Bigg[ {\rm Re} \left[C_9 C_{10}^*\right] \;  \omega_{910}^{\rm (em)}(\s) +
\as \kappa \; {\rm Re} \left[(C_2 + C_F C_1) C_{10}^* \;  \omega_{210}^{\rm (em)}(\s) \right]\Bigg]\Bigg\} \; \; ,
\label{deltaFBA}
\eea
with
\bea
\omega_{710}^{\rm (em)}(\s) & = & 
\ln \left(\frac{m_b^2}{m_\ell^2}\right)\,\left[ \frac{7 - 16\,\sqrt{\s} + 9\,\s}{
 4\,\left( 1 - \s \right)} + \ln (1 - \sqrt{\s}) + \frac{1+3 \,\s}{1-\s} \, \ln \!\left(\frac{1 + \sqrt{\s}}{2}\right)
- \frac{\s \,\ln \s}{\left( 1 - \s \right)} \right] \;,\nnb\\
&& \label{omega710em}\\
\omega_{910}^{\rm (em)}(\s) & = & 
\ln \left(\frac{m_b^2}{m_\ell^2}\right)\,\left[ -\frac{5 - 16\,\sqrt{\s} + 11\,\s}{
 4\,\left( 1 - \s \right)} + \ln (1 - \sqrt{\s}) \right.\nnb \\
 && \nnb\\
 && \hspace*{54pt}\left.+ \frac{1-5 \,\s}{1-\s} \, \ln \!\left(\frac{1 + \sqrt{\s}}{2}\right)
- \frac{(1-3 \,\s) \,\ln \s}{\left( 1 - \s \right)} \right] \;,\label{om910em}\\
 && \nnb\\  \label{omega210em}
\omega_{210}^{\rm (em)}(\s) & = & 
\ln \left(\frac{m_b^2}{m_\ell^2}\right)\,\left[-\frac{\Sigma_7(\s)+ i \,\Sigma_7^I(\s)}{24 \,\s \,(1-\s)^2}\right] + \frac{8}{9} \,
\omega_{910}^{\rm (em)}(\s)\,\ln\!\left(\frac{\mu_b}{5\gev}\right)\;.
\eea
%\bea \label{om710em}
%\omega_{710}^{\rm (em)}(\s) & = & 
%\ln \left(\frac{m_b^2}{m_\ell^2}\right)\,\left[ \frac{7 - 16\,\sqrt{\s} + 9\,\s}{
% 4\,\left( 1 - \s \right)} + \ln (1 - \sqrt{\s}) + \frac{1+3 \,\s}{1-\s} \, \ln \!\left(\frac{1 + \sqrt{\s}}{2}\right)
%- \frac{\s \,\ln \s}{\left( 1 - \s \right)} \right] \;,\nnb\\
%&& \\
%\omega_{910}^{\rm (em)}(\s) & = & 
%\ln \left(\frac{m_b^2}{m_\ell^2}\right)\,\left[ -\frac{5 - 16\,\sqrt{\s} + 11\,\s}{
% 4\,\left( 1 - \s \right)} + \ln (1 - \sqrt{\s}) \right.\nnb \\
% && \nnb\\
% && \hspace*{54pt}\left.+ \frac{1-5 \,\s}{1-\s} \, \ln \!\left(\frac{1 + \sqrt{\s}}{2}\right)
%- \frac{(1-3 \,\s) \,\ln \s}{\left( 1 - \s \right)} \right] \;,\\
% && \nnb\\
%\omega_{210}^{\rm (em)}(\s) & = & 
%\ln \left(\frac{m_b^2}{m_\ell^2}\right)\,\left[-\frac{\Sigma_7(\s)+ i \,\Sigma_7^I(\s)}{24 \,\s \,(1-\s)^2}\right] + \frac{8}{9} \,
%\omega_{910}^{\rm (em)}(\s)\,\ln\!\left(\frac{\mu_b}{5\gev}\right)\;.
%\eea
%
The functions $\omega_{710}^{\rm (em)}(\s)$ and $\omega_{910}^{\rm (em)}(\s)$ are known in the entire $q^2$-region, whereas the function $\omega_{210}^{\rm (em)}(\s)$ was obtained by a least squares fit. The  function $\Sigma_7$, valid in the low-$\s$-region, reads
\bea
\Sigma_7(\s) &=&  -0.259023 -28.424\, \s + 205.533\, \s^2 -603.219\, \s^3 + 722.031\, \s^4 \; ,\\
%\Sigma_7^I(\s) &=& (0.0452916 -4.29462\, \s + 44.7504\, \s^2 -123.953\, \s^3) \, \theta(\s-0.17) \; .
\Sigma_7^I(\s) &=& [-12.20658 - 215.8208 \, (\s-a) + 412.1207\, (\s-a)^2] \, (\s-a)^2\, \theta(\s-a) \; ,
\eea
with $a=(4 m_c^2/m_b^2)^2$. % \simeq 0.17066$.
In the high-$\s$-region the function $\Sigma_7$ reads ($\delta=1-\s$)
%
%\bea
%\Sigma_7(\s) &=&  -10.151 + 81.372\, \s -130.951\, \s^2 + 59.730\, \s^3 \; ,\\
%\Sigma_7^I(\s) &=& 30.629 -210.356\, \s + 546.335\, \s^2 -582.380\, \s^3+ 215.772\, \s^4  \; .
%\eea
\bea
\Sigma_7(\s) &=&  77.0256 \, \delta^2 - 264.705 \, \delta^3 + 595.814 \, \delta^4 - 610.1637 \, \delta^5 \; ,\\
\Sigma_7^I(\s) &=&  135.858\, \delta^2 - 618.990\, \delta^3 +1325.040\, \delta^4 - 1277.170\, \delta^5  \; .\label{sigma7}
\eea  
The polynomials in the high-$\s$-region were obtained such as to have a double zero at $\s=1$.

\subsection{Log-enhanced corrections to the BR, high-$q^2$ region}\label{sec:BRhighs}	
The log-enhanced corrections to the differential decay width given in Eq.~(\ref{bu}) read
\bea
\frac{{\rm d} \Delta \Gamma}{d\s} &=&
\frac{G_F^2 m_b^5}{48 \pi^3} |V_{tb} V_{ts}^{\ast}|^2 (1-\s)^2 \as \kappa %\ln \frac{m_b^2}{m_\ell^2} 
\Bigg\{ 8\, (1+ 2\s) \Bigg[ 
|C_9|^2 \;  \omega_{99}^{\rm (em)}(\s)  
+|C_{10}|^2 \; \omega_{1010}^{\rm (em)} (\s) 
\nnb\\
& &
+\as\kappa \; {\rm Re} \left[(C_2 + C_F C_1) C_9^* \;  \omega_{29}^{\rm (em)}(\s) \right]
+\as^2 \kappa^2 \; (C_2 + C_F C_1)^2 \;  \omega_{22}^{\rm (em)} (\s)
\Bigg]
\nnb \\
& & 
+96 \, \Bigg[ 
\as \kappa \; {\rm Re} \left[C_7 C_9^*\right] \; \omega_{79}^{\rm (em)}(\s) 
+\as^2\kappa^2 \; {\rm Re} \left[(C_2 + C_F C_1) C_7^* \;  \omega_{27}^{\rm (em)}(\s) \right]
\Bigg] 
\nnb\\
& &
+8\, (4+\frac{8}{\s})\as^2 \kappa^2 \; |C_{7}|^2 \; \omega_{77}^{\rm (em)} (\s) \Bigg\} \; \; .
\label{deltagamma}
\eea
Exact analytical expressions are known for the functions $\omega_{99}^{\rm (em)}(\s)$, $\omega_{1010}^{\rm (em)}(\s)$, $\omega_{77}^{\rm (em)}(\s)$, and $\omega_{79}^{\rm (em)}(\s)$ (see Eqs.~(94) and (100) --~(102) of Ref.~\cite{Huber:2005ig}). Therefore, they hold in the entire $q^2$-region, while the other $\omega$-functions are  obtained by a
least-squares fit in the high-$q^2$-region (for fixed values of $m_b$ and $m_c$):
\bea
\label{omega29} \omega_{29}^{\rm (em)}(\s) & = & 
\ln \left(\frac{m_b^2}{m_\ell^2}\right)\,\left[\frac{\Sigma_4(\s)+ i \,\Sigma_4^I(\s)}{8 (1-\s)^2 (1+2\s)}\right] + \frac{16}{9} \,
\omega_{1010}^{\rm (em)}(\s)\,\ln\!\left(\frac{\mu_b}{5\gev}\right)\;,\\
\omega_{22}^{\rm (em)}(\s) & = & 
\ln \left(\frac{m_b^2}{m_\ell^2}\right)\,\left[\frac{\Sigma_5(\s)}{8 (1-\s)^2 (1+2\s)} +  
\, \frac{\Sigma_4(\s)}{9 (1-\s)^2 (1+2\s)}\ln\!\left(\frac{\mu_b}{5\gev}\right)\right] \nnb\\ & & \nnb \\
    &&+ \, \frac{64}{81} \; \omega_{1010}^{\rm (em)}(\s)\, \ln^2\!\left(\frac{\mu_b}{5\gev}\right)
\;,\\ & & \nnb\\
\omega_{27}^{\rm (em)}(\s) & = & 
\ln \left(\frac{m_b^2}{m_\ell^2}\right)\,\left[\frac{\Sigma_6(\s)+ i \, \Sigma_6^I(\s)}{96 (1-\s)^2}\right] + \frac{8}{9} \,
\omega_{79}^{\rm (em)}(\s) \, \ln\!\left(\frac{\mu_b}{5\gev}\right)\; ,
\label{om79em}
\eea
with the  functions $\Sigma_i$, valid in the high-$\s$-region ($\delta = 1-\s$):
\bea
\Sigma_4(\s) &=&  -148.061 \, \delta^2 + 492.539 \, \delta^3 - 1163.847 \, \delta^4 +  1189.528 \, \delta^5 \; ,\\
\Sigma_4^I(\s) &=& - 261.287 \, \delta^2 + 1170.856 \, \delta^3 - 2546.948 \, \delta^4 + 2540.023\, \delta^5 \; ,\\%
\Sigma_5(\s) &=& - 221.904\, \delta^2 + 900.822\, \delta^3 -2031.620 \, \delta^4 + 1984.303\, \delta^5 \; ,\\ %
\Sigma_6(\s) &=&  -298.730\, \delta^2 + 828.0675\, \delta^3 -  2217.6355\, \delta^4 +  2241.792\,  \delta^5\; ,\\ %
\Sigma_6^I(\s) &=&  - 528.759\, \delta^2 +  2095.723\, \delta^3 - 4681.843\, \delta^4 + 5036.677\, \delta^5\; . \label{sigma6}
%\eea
%\bea
%\Sigma_4(\s) &=&  -2.851   - 88.341 \, \s + 182.603\, \s^2 -  91.411\, \s^3 \; ,\\
%\Sigma_4^I(\s) &=& -24.452 + 230.651\, \s - 733.446\, \s^2 + 869.312\, \s^3 - 342.065\, \s^4 \; ,\\%
%\Sigma_5(\s) &=&  -90.993  + 416.716\, \s - 868.541\, \s^2 + 848.182\, \s^3 - 305.364\, \s^4 \; ,\\ %
%\Sigma_6(\s) &=&  -138.223 + 215.226\, \s -  20.933\, \s^2 -  56.070\, \s^3\; ,\\ % 
%\Sigma_6^I(\s) &=&  \label{sigma6} + 81.070 - 305.884\, \s +  46.727\, \s^2 + 493.155\, \s^3 - 315.068\, \s^4\; .  \label{sigma6}
\eea  
The fits are excellent for $\s>0.65$. Again, the polynomials were obtained such as to have a double zero at $\s=1$.
%{\bf [[For the fits, I took $m_c=1.504 \; \gev$ and $m_b=4.68 \; \gev$ as in the previous paper.]]}

\subsection{Phenomenological implications of the collinear logarithms}
%\subsection{Collinear logarithm in the $ee$ and $\mu\mu$ final state}
After inclusion of the NLO QED matrix elements calculated in the previous
subsections the electron and muon channels receive different contributions due to terms involving $\ln(m_b^2/m_\ell^2)$. 
We emphasize that this is the only source of the difference between
the electron and muon channel.

However, as already pointed out in Ref.~\cite{Huber:2005ig}, the presence of this
logarithm is strictly related to the definition of the dilepton invariant mass. The collinear logarithm $\ln(m_b^2/m_\ell^2)$ 
disappears if all photons emitted by the final state on-shell leptons
are included in the definition of $q^2$: $(p_{\ell_1}+p_{\ell_2})^2
\rightarrow (p_{\ell_1}+p_{\ell_2}+p_\gamma)^2$. If none of these photons was included in the definition of $q^2$ -- i.~e.\ if a perfect separation of leptons and collinear photons was possible experimentally -- then our
  expressions containing $\ln(m_b^2/m_\ell^2)$ were directly applicable. This is the case for muons since the separation of muons
  and collinear photons is practically perfect~\cite{private}. For electrons with the current Babar and Belle setups the lepton mass gets replaced by an effective mass parameter $\Lambda$ which is found to be of the same
order as $m_\mu$. Hence our numerical results in section~\ref{sec:results} for muonic final states should be applied also to the case of electronic final states in the present Babar and Belle setups. All results in
section~\ref{sec:results} for the electron channel are given under the assumptions of perfect separation of electrons and collinear photons. We refer the reader to section 6 of Ref.~\cite{Huber:2005ig} for more details.

\subsection{Two-loop matrixelement of $P_9$}  
In Ref.~\cite{Gambinonew}, an estimate of the two-loop (NNLO) matrix element of the  operator $P_9$ --  denoted $\omega_{99}^{(2)}(\s)$ in Eq.~(74) of Ref.~\cite{Huber:2005ig} --  was presented which was neglected in all previous analyses. The estimate is based on the fact that the QCD corrections to $b \rightarrow s \ell^+\ell^-$ are identical to those of $b \rightarrow u\ell\nu\, (t \rightarrow b \ell\nu)$, in the limit of vanishing strange (bottom) quark mass. However, the specific two-loop calculation of the $b \rightarrow u\ell\nu$  decay was available as an expansion in $(1-\s)$~\cite{semi1}, while the two-loop contribution  of the top quark decay was only known as an expansion of $M_W^2 / m_t^2$ (which translates into an expansion in $q^2 / m_b^2$ for $b \rightarrow s \ell\ell$) up to the second order based on Pade approximation methods~\cite{semi2}. In the meanwhile, the QCD corrections to $b \rightarrow u\ell\nu$ are also known as an expansion in $\s$ \cite{semi3,semi4}.

Thus, in the low-$\s$ region we approximate the function $\omega_{99}^{(2)}$ by the in $\s =  \omega$ expanded  results
\be
\omega_{99}^{(2)}(\s)  \equiv X_2(\omega = \s) / X_0(\omega = \s) \label{omega99}
\ee
where $X_2(\omega)$ is given in Eq.~(60) of Ref.~\cite{semi3} and $X_0(\omega)$ in Eq.~(2) of Ref~\cite{semi4}. 

Analogously, in the high-$\s$ region we approximate the function $\omega_{99}^{(2)}$  by the in $(1-\s) = \delta$ expanded results of $X_2(\delta = 1-\s)$ and  $X_0(\delta = 1- \s)$ as  given in Eq.~(2) and (3) of Ref.~\cite{semi1}. Note that the normalization of the $X_i$ differs by a factor of 2 in Ref.~\cite{semi1} and Refs.~\cite{semi3,semi4}.\footnote{There is a typo in Eq.~(4) of Ref.~\cite{semi1}: the powers of the last four $\delta$'s have to be increased by 1 each.}

\subsection{Some nonperturbative subleties}

For a detailed discussion of nonperturbative corrections we refer the reader to Section 5 of Ref.~\cite{Adrian1}. Here we add few remarks on some specific issues.

The long-distance corrections related to the $c \bar c$ intermediate states originate from the non-per\-tur\-ba\-tive interactions of
the $c\bar c$ pair in the process $\bar B\to X_s c\bar c \to X_s \ell^+ \ell^-$. If the dilepton invariant mass is near the first two
$J^{PC}=1^{--}$ $c\bar c$ resonances ($\Psi$ and $\Psi'$), this effect is very large  and shows up as a peak in the dilepton mass
spectrum which  can easily be eliminated by suitable kinematical cuts. More delicate is the estimate of such long-distance effects away from the resonance peaks in the perturbative windows. 

Within the KS--approach~\cite{KS}, one reabsorbs charm rescattering effects into the matrix element of $P_9$; the effects of $b\to
c\bar{c} s$ operators is estimated by means of experimental data on $\sigma(e^+e^-\to c\bar c$ hadrons) using a dispersion relation. To
be more specific, the function $h(z,s)$ appearing in Eq.~(A.5) of Ref.~\cite{Adrian1} (which corresponds to  the expression $g(y_c) +
8/9 \log(m_b/m_c) - 4/9$ in Eq.~(72) of Ref.~\cite{Huber:2005ig}) is replaced by ($z=m_c^2/m_b^2$)
\begin{equation}\label{cbarc}
h(z,\s) \longrightarrow  h(z,0) + {s\over 3} P \int_{m_\pi^2}^{\infty}
d\s' \frac{ R^{c\bar{c}}_{\rm had} (\s') }{\s'(\s'-\s)}
+ i\frac{\pi}{3} R^{c\bar{c}}_{\rm had} (\s)~,
\end{equation}
where $R^{c\bar{c}}_{\rm had} (\s) = \sigma(e^+e^-\to c\bar c)/\sigma(e^+e^-\to \mu^+\mu^-)$.
This method is exact only in the limit in which the $\bar B\to X_s c\bar c$ transition can be factorized into the product of $\bar{s}b$
and $\bar{c}c$ colour-singlet currents. It is possible to take into account non-factorizable effects and reproduce the correct hadronic
branching ratios by multiplying $R^{c\bar{c}}_{\rm had}$ by a purely phenomenological factor $\kappa = 2.6$. However, a
model-independent way to estimate the non-factorizable $c\bar c$ long-distance effects far from the resonance region exists by means of
an expansion in inverse powers of the charm-quark mass~\cite{Buchalla:1997ky,Savagenew}. Having included those $1/m_c$ corrections, an
`inflation' of the factorizable KS-corrections using the phenomenological factor $\kappa$ would lead to a {\it double-counting}. The
numerical impact of the KS contribution to the integrated rate in the low- and high-s regions is about $+2$\% and $-10$\%, respectively.

In Section 5 of Ref.~\cite{Adrian1}, another large  nonperturbative error for the high-$q^2$ region  was located in the {\it linear} $1/m_b$ corrections. The physical observables defined in terms of a $q^2$-cut are sensitive to a $1/m_b$ term via the the uncertainty on the value of $m_{b,{\rm pole}}$ or equivalently via the relation of the pole mass and the hadron mass 
\be
M_B = m_{b} (1 + \bar \Lambda / m_b + O(1/m_b^2) ).  
\ee
Choosing as reference cut the value $\s_{\rm min} = 0.6$, the integrated normalized dilepton mass spectrum defined in terms of $q^2_{\rm min}$ can be written as 
\bea
R_{\rm cut}(q^2_{\rm min})
 &=& \int_{ q^2 > q^2_{\rm min} } d q^2 
\frac{ d \Gamma(\bar B\to X_s \ell^+\ell^-)}{\Gamma(\bar B\to X_c e \nu)} \nonumber\\
 &=&  \left\{ 1- 6.2 \left(\frac{q^2_{\rm min}}{m_b^2}-0.6 \right) + 
O \left[ \left(\frac{q^2_{\rm min}}{m_b^2}-0.6 \right)^2 \right] \right\}\times \int_{0.6}^{1} ds~R(s)~,
\eea
which implies 
\be
\frac{\delta R_{\rm cut}}{ R_{\rm cut}} \approx 7.4 \frac{\delta m_b }{m_b}~.
\ee
Using the pole mass scheme with $\delta m_b=0.1$~GeV, this leads to a $\approx 15\%$  error on $R_{\rm cut}$. However, this error gets now significantly reduced in our updated analysis using the kinematical
1S scheme for the $m_b$ mass.
\begin{table}[t]
\begin{center}
\begin{displaymath}
\begin{tabular}{|l|l|}
\hline\spp 
$\alpha_s (M_z) = 0.1189 \pm 0.0010$~\cite{Bethke:2006ac} &  
  $m_e = 0.51099892 \;\mev $ \\ \spp 
$\alpha_e (M_z) =  1/127.918 $ & 
  $m_\mu = 105.658369 \;\mev$ \\ \spp 
$s_W^2 \equiv \sin^2\theta_W = 0.2312$ & 
  $m_\tau = 1.77699 \;\gev$ \\ \spp 
$|V_{ts} V_{tb}/V_{cb}|^2 = 0.962 \pm 0.002$~\cite{Charles:2004jd} &
  $m_c(m_c) = (1.224 \pm 0.017 \pm 0.054)\;\gev$~\cite{Hoang:2005zw}\\\spp 
$|V_{ts} V_{tb}/V_{ub}|^2 = (1.28 \pm 0.12 ) \times 10^2$~\cite{Charles:2004jd} &
  $m_b^{1S} = (4.68 \pm 0.03)\;\gev$~\cite{Bauer:2004ve} \\ \spp 
$BR(B\to X_c e \bar\nu)_{\rm exp}=0.1061 \pm 0.0017$~\cite{Aubert:2004aw} & 
  $m_{t,{\rm pole}}= (170.9 \pm 1.8) \;\gev$~\cite{topmass}\\ \spp 
$M_Z = 91.1876\;\gev$ & 
  $m_B = 5.2794\;\gev$ \\ \spp 
$M_W = 80.426\;\gev$ & 
  $C = 0.58 \pm 0.01$~\cite{Bauer:2004ve} \\ \spp 
$\lambda_2^{\rm eff} = (0.12 \pm 0.02)\;\gev^2$ & 
  $\rho_1 = (0.06 \pm 0.06)\;\gev^3$~\cite{Bauer:2004ve} \\ \spp
$\lambda_1^{\rm eff} = (-0.243 \pm 0.055)\;\gev^2$~\cite{Hoang:2005zw} & 
  $f_u^0+f_s = (0 \pm 0.2)\;\gev^3$~\cite{Ligeti:2007sn} \\ \spp
$f_u^0-f_s = (0 \pm 0.04)\;\gev^3$~\cite{Ligeti:2007sn} &
$f_u^\pm = (0 \pm 0.4)\;\gev^3$~\cite{Ligeti:2007sn} \\ \hline
\end{tabular}
\end{displaymath}
\caption{Numerical inputs that we use in the phenomenological analysis. Unless explicitly specified, they are taken from PDG 2004~\cite{Eidelman:2004wy}.}
\label{tab:inputs}
\end{center}
\end{table}
\section{Numerical results}
\label{sec:results}

The numerical inputs that we adopt are summarized in Table~\ref{tab:inputs}.
\subsection{Branching ratio in the high-$q^2$ region}
For the branching ratio integrated over the region $q^2 > 14.4\; \gev^2$  we find:
\bea
{\cal B}_{\mu\mu}^{\rm high} & = & \label{muonBR} 
2.40 \times 10^{-7} \; \Big(
1
+ \left[ {}^{+0.01}_{-0.02} \right]_{\mu_0}
+ \left[ {}^{+0.14}_{-0.06} \right]_{\mu_b}
\pm 0.02_{m_t} 
+ \left[ {}^{+0.006}_{-0.003} \right]_{{C,m_c}}
\pm 0.05_{m_b}
+ \left[ {}^{+0.0002}_{-0.001} \right]_{\alpha_s}
   \nnb \\ & &
\pm 0.002_{\rm CKM} 
\pm 0.02_{{\rm BR}_{sl}} 
\pm 0.05_{\lambda_2}
\pm 0.19_{\rho_1}
\pm 0.14_{f_s}
\pm 0.02_{f_u} 
\Big)  \nnb \\
&=& 2.40 \times 10^{-7} \; (1^{+0.29}_{-0.26} ) \; ,\\
{\cal B}_{ee}^{\rm high} & = & \label{electronBR} 
2.09 \times 10^{-7} \; \Big(
1
+ \left[ {}^{+0.02}_{-0.04} \right]_{\mu_0}
+ \left[ {}^{+0.16}_{-0.08} \right]_{\mu_b}
\pm 0.02_{m_t} 
+ \left[ {}^{+0.005}_{-0.0009} \right]_{{C,m_c}}
\pm 0.05_{m_b}
+ \left[ {}^{+0.0003}_{-0.002} \right]_{\alpha_s}
   \nnb \\ & &
\pm 0.002_{\rm CKM} 
\pm 0.02_{{\rm BR}_{sl}} 
\pm 0.05_{\lambda_2}
\pm 0.22_{\rho_1}
\pm 0.16_{f_s}
\pm 0.02_{f_u} 
\Big)  \nnb \\
&=& 2.09 \times 10^{-7} \; (1^{+0.32}_{-0.30} ) \; .
\eea
The scale uncertainty has been estimated by varying the matching scale $\mu_{0}$ and the low-energy scale $\mu_b$ by factors of 2 around their central values (120 and 5 GeV, respectively). In our approach we normalized the $b\to s\ell\ell$ decay width to the semileptonic rate averaged over neutral and charged B mesons; hence, we average $f_u^0$ and $f_u^\pm$. From the values quoted in Table~\ref{tab:inputs} we obtain: $f_u= 0\pm 0.2$ and $f_s = 0 \pm 0.1$~\footnote{Regarding the weak annihilation contributions  ($f_u^{0}$, $f_u^{+}$, $f_s^{0}$, $f_s^{+}$) we follow the analysis in Ref.~\cite{Ligeti:2007sn}. Here are some additional remarks in order: Using flavor $SU(3)$ one concludes 
that $f_u^0 \sim f_s^0$. However, we stress that  the additional assumption 
$f_s^0 \sim f_s^+$  comes from neglecting iso--singlet effects in 
the $SU(3)$ decomposition of $\bar b s \; \bar s b$. 
Nevertheless, due to a mismatch in the flavor of the spectator quark 
in the $B$ meson with the quark appearing
in the operators ( $\bar b u \; \bar u b$ and $\bar b s \; \bar s b$), it is not unreasonable to assume $f_s^+ \; , f_u^0 < f_u^+$. 
This leads to the relation $f_u^0 \sim f_s^0 \sim f_s^+ < f_u^+$~\cite{Ligeti:2007sn} that we use in the numerics.  
We note again that in the definition of $\cal R$ we use only neutral semileptonic decays, thus, the dependence on $f_u^{+}$ disappears in the ratio. 
The actual numerical inputs we adopt are taken from Ref.~\cite{Voloshin:2001xi, Ligeti:2007sn}. Because all our expressions for the branching ratio and 
the ratio ${\cal R}_0$ are linear
in these non-perturbative parameters, hence it is trivial to adjust our results to accommodate for 
different sets of non-perturbative inputs.
}

The other parametric uncertainties are obtained by varying the inputs within the errors given in Table~\ref{tab:inputs}. We assume the errors on C and $m_c$ to be fully correlated. The total error is obtained by adding the individual uncertainties in quadrature. We note that here and in the following all errors are parametric or  perturbative  uncertainties only and
that subleading nonperturbative corrections of order $O(\alpha_s \Lambda/m_b)$ will give an additional uncertainty.

\begin{figure}[t]
\begin{center}
\includegraphics[scale=.8]{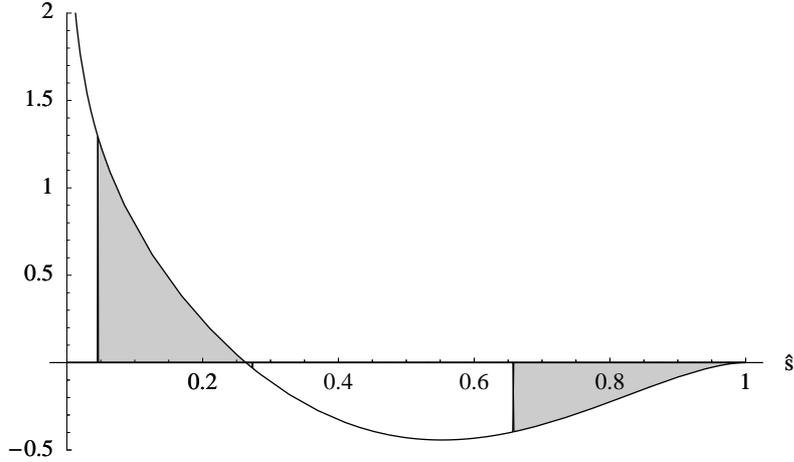}
\end{center}
\caption{The log--enhanced part of $(1-\s)^2 (1+2\s) \, \omega^{\rm (em)}_{99}(\s)$, see Eq.~(94) of Ref.~\cite{Huber:2005ig}. This function integrates to zero, but gives contributions of opposite sign to the low-$\s$ and high-$\s$
branching ratio, respectively (shaded areas). The moduli of the absolute shifts are of comparable size. However, the relative effect is much larger in the high-$\s$ compared
to the low-$\s$ region due to the steep decrease of the differential decay width at large $\s$.}
\label{fig:om99plot}
\end{figure}

Log--enhanced QED bremsstrahlung corrections shift the central value by about $-8$\% and $-20$\% for the muonic and the electronic final state, respectively. These relative shifts are much larger than the corresponding ones in the
low-$q^2$ region. In fact, these corrections vanish when integrated over the whole dilepton invariant mass spectrum, but are relevant if one is restricted to certain regions in phase space. One observes that the absolute shifts in the
central values of the branching ratio are of similar size but of opposite sign in the low-$q^2$ compared to the high-$q^2$ region, namely for the muonic final state $+3 \times 10^{-8}$ and $-2 \times 10^{-8}$ respectively. But since
the differential decay width decreases steeply at large $q^2$, the relative effect is much more pronounced in the high-$q^2$ compared to the low-$q^2$ region. See also Fig.~\ref{fig:om99plot} for illustration.

\subsection{Ratio ${\cal R}(s_0)$}\label{sec:R0res}
The numerical results we obtain for the ratio ${\cal R}(s_0)$ discussed in Section~\ref{sec:R0} are for $s_0=14.4$ $\gev^2$:
\bea
{\cal R}(s_0)_{\mu\mu}^{\rm high} & = & \label{muonR} 
2.29 \times 10^{-3} \Big(
1
\pm 0.04_{\rm scale}
\pm 0.02_{m_t} 
\pm 0.01_{{C,m_c}}
\pm 0.006_{m_b}
\pm 0.005_{\alpha_s}
\pm 0.09_{\rm CKM} 
   \nnb \\ & &
\pm 0.003_{\lambda_2}
\pm 0.05_{\rho_1}
\pm 0.03_{f_u^0+f_s}
\pm 0.05_{f_u^0-f_s} 
\Big) \nnb \\
&=& 2.29 \times 10^{-3} ( 1 \pm 0.13) \; ,\\
{\cal R}(s_0)_{ee}^{\rm high} & = & \label{electronR} 
1.94 \times 10^{-3} \Big(
1
\pm 0.06_{\rm scale}
\pm 0.02_{m_t} 
\pm 0.02_{{C,m_c}}
\pm 0.004_{m_b}
\pm 0.006_{\alpha_s}
\pm 0.09_{\rm CKM} 
   \nnb \\ & &
\pm 0.01_{\lambda_2}
\pm 0.09_{\rho_1}
\pm 0.05_{f_u^0+f_s}
\pm 0.05_{f_u^0-f_s} 
\Big) \nnb \\
&=& 1.94 \times 10^{-3} (1 \pm 0.16) \; .
\eea
Note that uncertainties from poorly known $O(1/m_b^3)$ power corrections are now under control; the largest source of error is $V_{ub}$.

\subsection{Branching ratio in the low-$q^2$ region}

For the branching ratio integrated in the range
$1\;\gev^2 < m^2_{\ell\ell} < 6\;\gev^2$ it was found in~Ref.~\cite{Huber:2005ig}:
\bea
\hskip -0.5cm
{\cal B}_{\mu\mu} & = & \label{muonBRlow} 
\Big[ 
 1.59 
\pm 0.08_{\rm scale}
\pm 0.06_{m_t} 
\pm 0.024_{ C,m_c }
\pm 0.015_{m_b} \nnb \\
&& \hspace*{28.5pt} \pm 0.02_{\alpha_s(M_Z)}
\pm 0.015_{\rm CKM} 
\pm 0.026_{{\rm BR}_{sl}} 
\Big] \times 10^{-6} = (  1.59  \pm 0.11 ) \times 10^{-6} \;, \\
\hskip -0.5cm
{\cal B}_{ee} & = & \label{electronBRlow} 
\Big[ 
 1.64  
\pm 0.08_{\rm scale} 
\pm 0.06_{m_t} 
\pm 0.025_{ C,m_c }
\pm 0.015_{m_b} \nnb \\
&& \hspace*{28.5pt} \pm 0.02_{\alpha_s(M_Z)}
\pm 0.015_{\rm CKM}
\pm 0.026_{{\rm BR}_{sl}}
\Big] \times 10^{-6}  = (  1.64   \pm 0.11) \times 10^{-6} \;.
\eea 
We note that we find for ${\cal B}_{\mu\mu} $ a 
$+1.8\%$ shift of the central  value due to the KS-corrections which were discussed in Section 3.4 and 
which were not included in the previous analysis. Moreover, the update of the input parameters (CKM, $m_t, \alpha_s$)  leads to a $-3.1\%$ shift; thus, we end up to an overall change of $-1.3\%$  within our new  phenomenolgical
analysis compared with the previous one in Ref.~\cite{Huber:2005ig}. The total error, being from parametric and perturbative
uncertainties only, remains unchanged.

\subsection{Integrated FBA in low-$s$ region}
By the end of the current $B$ factories the fully differential FBA will not be accessible experimentally, contrary to the integral over
one or more bins in the low-$q^2$ region. However, these integrals can already serve of gain important information on the shape of the
FBA and to constrain the parameter space of new physics models (see for instance~\cite{AGHL}).
We now give the results for the integrated FBA based on Eq.~(\ref{doubleratio}). We subdivide the low-$\s$ region into the two bins $s
\in [1,3.5]$~$\gev^2$ and $s \in [3.5,6]$~$\gev^2$, which we will call bin 1 and bin 2, respectively\footnote{Predictions for different
bins can be produced upon request.}. We then integrate the numerator of the double ratio, Eq.~(\ref{fba}), over the respective bin, and
divide the result by the denominator of the double ratio, Eq.~(\ref{br}) integrated over the same bin, i.~e.\ we first integrate FBA and
branching ratio separately and subsequently divide the two numbers. Our results read:
\bea
\left(\bar{\cal A}_{\mu\mu}^{FB}\right)_{{\rm bin} 1} & = &
 \Big[ \, -9.09
\pm 0.83_{\rm scale}
\pm 0.03_{m_t} 
\pm 0.24_{m_c,C}
\pm 0.20_{m_b} 
\pm 0.18_{\alpha_s(M_Z)}
\pm 0.02_{\lambda_2} \, \Big] \, \%  \nnb \\
& = & \label{muonbin1} 
\Big[ \, -9.09 \pm 0.91 \, \Big] \, \% \;, \\
\hskip -0.5cm
\left(\bar{\cal A}_{ee}^{FB}\right)_{{\rm bin} 1} & = & 
 \Big[ \, -8.14 
\pm 0.80_{\rm scale} 
\pm 0.03_{m_t} 
\pm 0.23_{m_c,C}
\pm 0.19_{m_b} 
\pm 0.18_{\alpha_s(M_Z)} 
\pm 0.02_{\lambda_2}\, \Big] \, \%  \nnb \\
& = & \label{electronbin1} 
\Big[ \, -8.14 \pm 0.87 \, \Big] \, \% \;, \\
\left(\bar{\cal A}_{\mu\mu}^{FB}\right)_{{\rm bin} 2} & = & 
 \Big[ \, +7.80
\pm 0.55_{\rm scale}
\pm 0.02_{m_t} 
\pm 0.31_{m_c,C}
\pm 0.34_{m_b} 
\pm 0.16_{\alpha_s(M_Z)} 
\pm 0.20_{\lambda_2}\, \Big] \, \% \nnb \\
& = & \label{muonbin2} 
\Big[ \, +7.80 \pm 0.76 \, \Big] \, \% \;, \\
\hskip -0.5cm
\left(\bar{\cal A}_{ee}^{FB}\right)_{{\rm bin} 2} & = &  
 \Big[ \, +8.27
\pm 0.47_{\rm scale} 
\pm 0.02_{m_t} 
\pm 0.30_{m_c,C}
\pm 0.33_{m_b} 
\pm 0.15_{\alpha_s(M_Z)} 
\pm 0.19_{\lambda_2}\, \Big] \, \%  \nnb \\
& = & \label{electronbin2} 
\Big[ \, +8.27 \pm 0.69 \, \Big] \, \% \;.
\eea
The total error is obtained by adding the individual ones in quadrature. For the entire low-$\s$ region, we get 
\bea
\hskip -0.5cm
\left(\bar{\cal A}_{\mu\mu}^{FB} \right)_{\rm low} & = &
 \Big[ \, -1.50
\pm 0.78_{\rm scale}
\pm 0.02_{m_t} 
\pm 0.29_{m_c,C}
\pm 0.27_{m_b} 
\pm 0.18_{\alpha_s(M_Z)} 
\pm 0.10_{\lambda_2}\, \Big] \, \% \nnb \\
& = & \label{muonentire} 
\Big[ \, -1.50 \pm 0.90 \, \Big] \, \% \;, \\
\hskip -0.5cm
\left( \bar{\cal A}_{ee}^{FB} \right)_{\rm low}& = & 
 \Big[ \, -0.86
\pm 0.73_{\rm scale} 
\pm 0.01_{m_t} 
\pm 0.28_{m_c,C}
\pm 0.26_{m_b} 
\pm 0.18_{\alpha_s(M_Z)} 
\pm 0.10_{\lambda_2}\, \Big] \, \% \nnb \\
& = & \label{electronentire} 
\Big[ \, -0.86 \pm 0.85 \, \Big] \,\% \;.
\eea

The relative errors in the respective bins are considerably smaller than for the entire low-$\s$ region since the respective values
in each bin are similar in size and of opposite sign.

\subsection{Analysis of the zero of the FBA}\label{sec:analysezeroFBA}
The basic formula for the extraction of the zero is Eq.~(\ref{doubleratio}). We expand the two ratios separately in $\alpha_s$,
$\kappa$, $\lambda_1$ and $\lambda_2$, and keep all the  terms as specified in section~\ref{perturbative}. It is understood that
also the conversion of the mass scheme for the botton, the charm and the top quark is performed in the way described there. 
The results for $q_0^2$, the zero of the FBA in the low-$s$ region, are 
\bea
(q_0^2)_{\mu\mu} & = & 
 \Big[ \, 3.50
\pm 0.10_{\rm scale}
\pm 0.002_{m_t} 
\pm 0.04_{m_c,C}\nnb \\
&& \label{muonzero} \hspace*{30pt}\pm 0.05_{m_b} 
\pm 0.03_{\alpha_s(M_Z)} 
\pm 0.001_{\lambda_1}
\pm 0.01_{\lambda_2} \, \Big] \, \gev^2 = ( 3.50 \pm 0.12) \, \gev^2\; , \nnb \\
& & \\
(q_0^2)_{ee} & = & 
 \Big[ \, 3.38
\pm 0.09_{\rm scale} 
\pm 0.002_{m_t} 
\pm 0.04_{m_c,C} \nnb \\
 && \label{electronzero} \hspace*{30pt} \pm \, 0.04_{m_b} 
\pm 0.03_{\alpha_s(M_Z)} 
\pm 0.002_{\lambda_1}
\pm 0.01_{\lambda_2}\, \Big] \, \gev^2 = ( 3.38 \pm 0.11 ) \, \gev^2 \; .\nnb \\
\eea
The central values are obtained for the matching scale $\mu_0 = 120\;\gev$ and the low-energy scale $\mu_b=5\;\gev$. The uncertainty
from missing higher order perturbative corrections have been estimated by increasing and decreasing the scales $\mu_{0}$ and $\mu_b$ by
factors of 2. Uncertainties induced by $m_t$, $m_b$, $m_c$, $C$, $\alpha_s(M_Z)$, $\lambda_{1}$ and $\lambda_2$ are obtained by varying
the various inputs within the errors given in Table~\ref{tab:inputs}. We assume the errors on C and $m_c$
to be fully correlated. The total errors are again obtained by adding the individual ones in quadrature. In order to show the stability
of the zero under change of the $b$ quark mass scheme we collect in Table~\ref{tab:schemes} the results we obtained in the 1S,
$\overline{\rm MS}$ and pole schemes.
\begin{table}[t]
\begin{center}
\begin{tabular}{|l|lll|}
\hline
\spp        & 1S     & $\overline{\rm MS}$ & pole \\ \hline
\spp  $\mu$ & 3.50   & 3.47                & 3.52 \\
\spp  $e$   & 3.38   & 3.34                & 3.41 \\ \hline
\end{tabular}
\caption{Dependence of the zero of the FBA on the $b$ quark mass scheme. The input values are $\overline{m}_b(\overline{m}_b)=4.205 \,
\gev$~\cite{Boughezal:2006px} and $m_{b, \rm pole}=4.8 \, \gev$.
\label{tab:schemes}}
\end{center} 
\end{table}

The total errors are 3.4\% and 3.3\% respectively and therefore quite small. It is often argued that especially the small $\mu$
dependence at the zero is an accident and should  be increased by hand. We argue in the following that the small $\mu$ dependence is a
reasonable reflection of the perturbative error.

\begin{figure}[t]
\begin{center}
\includegraphics{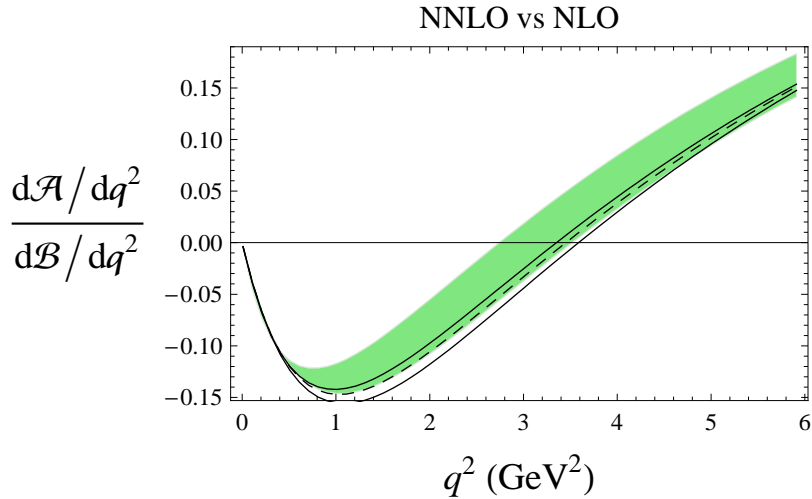}
\end{center}
\caption{$\mu_b$-dependence of the forward backward asymmetry for the muonic final state. The lines are the NNLO QCD result; the
dashed line corresponds to $\mu_b = 5$~$\gev$, and the solid lines to $\mu_b = 2.5,10$~$\gev$. The shaded area is the region spanned by
the NLO asymmetry for $2.5\; \gev<\mu_b< 10\; \gev$.}
\label{fig:AFBlowsplot}
\end{figure}

One test of our estimation of the perturbative error on the zero consists in comparing the central
values and $\mu_b$ dependences of lower--order predictions. 
The comparison between the NNLO+QED, NNLO and NLO extraction of the zero reads (here we consider only the muon channel and quote only
the scale uncertainty):
\bea
q_0^2 & = & \cases{
(3.50 \pm 0.10) \; \gev^2  & NNLO + QED \cr
(3.45 \pm 0.11 ) \; \gev^2  & NNLO \cr
(3.11 \pm 0.39 ) \; \gev^2  & NLO \, . \cr}
\label{zero-comp}
\eea
In Fig.~\ref{fig:AFBlowsplot}, we plot the $\mu_b$ dependence of the FBA in the low-$q^2$ region and compare NNLO and NLO QCD results.
Here we note that the scale uncertainty is maximal at $q^2 \sim 1\; \gev^2$, decreases smoothly at larger $q^2$ and almost vanishes at
the edge of the low-$q^2$ region. Moreover, the band between the two solid lines, representing the scale uncertainty of the NNLO FBA,
lies to a large extent within the shaded area of the NLO scale uncertainty. The same holds true if we compared NNLO+QED vs. NLO QCD
results.

\begin{figure}[t]
\begin{center}
\includegraphics{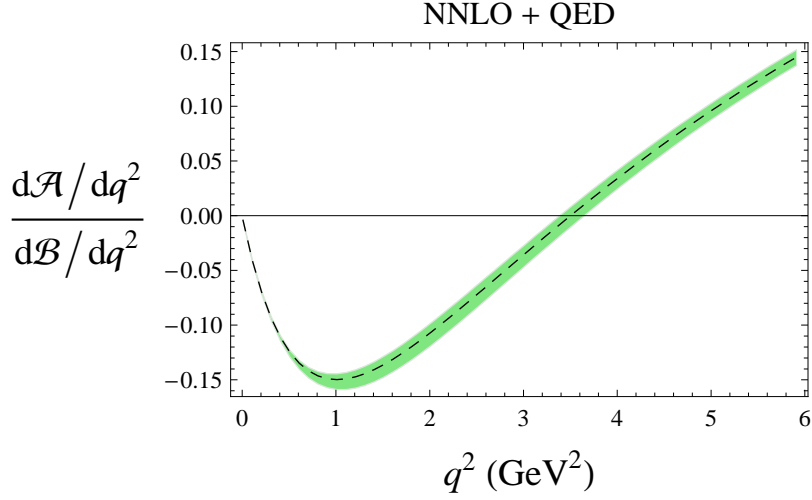}
\end{center}
\caption{The full NNLO+QED asymmetry (dashed line) and the total -- parametric and perturbative -- error band (shaded area).}
\label{fig:AFBlowsploterror}
\end{figure}

In Fig.~\ref{fig:AFBlowsploterror} we show the entire -- parametric and perturbative -- error band of the full NNLO+QED asymmetry over
the whole low-$s$ region. The plot shows that the perturbative expansion
converges nicely everywhere in the low-$q^2$ region.

The numerical results of Eq.~(\ref{zero-comp}) and the plots therefore suggest that
the variation of the scale $\mu_b$ properly describes the size of the missing perturbative corrections.

As already stressed before, the errors considered here are parametric and perturbative ones only; and unknown subleading
nonperturbative corrections of order $O(\alpha_s \Lambda/m_b)$ may give an additional uncertainty of order $5\%$.


\subsection{New-physics formulae}
Here, we present numerical formulae for the various observables for non-SM values of the high-scale Wilson coefficients of the operators
$P_7$, $P_8$, $P_9$ and $P_{10}$:
\bea\label{numform} 
{\cal B}_{\mu\mu}^{\rm high} & = & \Big[\; 
2.399- 0.002576 \; {\cal I} (\delta R_{10})- 0.01277 \; {\cal I} (\delta R_{7})+ 0.0002656 \; {\cal I} ( \delta R_7 \delta R_8^*)
\nonumber\\ & & 
+ 0.0004108 \; {\cal I} ( \delta R_7 \delta R_9^*)
+ 0.002027 \; {\cal I} (\delta R_{8})- 0.00003375 \; {\cal I} ( \delta R_8 \delta R_{10}^*)
\nonumber\\ & & 
+ 0.001676 \; {\cal I} ( \delta R_8 \delta R_9^*)+ 0.1079 \; {\cal I} (\delta R_{9})
+ 3.022 \; {\cal R} (\delta R_{10})+ 0.001146 \; {\cal R} ( \delta R_{10} \delta R_7^*)
\nonumber\\ & & 
+ 0.0001236 \; {\cal R} ( \delta R_{10} \delta R_8^*)- 0.0173 \; {\cal R} ( \delta R_{10} \delta R_9^*)- 0.1312 \; {\cal R} (\delta R_{7})- 0.0143 \; {\cal R} (\delta R_{8})
\nonumber\\ & & 
+ 0.0008111 \; {\cal R} ( \delta R_8 \delta R_7^*)+ 0.9375 \; {\cal R} (\delta R_{9})- 0.0568 \; {\cal R} ( \delta R_9 \delta R_7^*)
\nonumber\\ & & 
- 0.006099 \; {\cal R} ( \delta R_9 \delta R_8^*)
+ 1.558 \; |\delta R_{10}|^2+ 0.003436 \; |\delta R_7|^2+ 0.00004162 \; |\delta R_8|^2
\nonumber\\ & & 
+ 0.2231 \; |\delta R_9|^2
\; \Big] \times 10^{-7} \; , \\
{\cal B}_{ee}^{\rm high} & = & \Big[\; 
 2.085- 0.002576 \; {\cal I} (\delta R_{10})- 0.011 \; {\cal I} (\delta R_{7})+ 0.0002656 \; {\cal I} ( \delta R_7 \delta R_8^*)
\nonumber\\ & & 
+ 0.0004108 \; {\cal I} ( \delta R_7 \delta R_9^*)+ 0.002162 \; {\cal I} (\delta R_{8})- 0.00003375 \; {\cal I} ( \delta R_8 \delta R_{10}^*)
\nonumber\\ & & 
+ 0.001676 \; {\cal I} ( \delta R_8 \delta R_9^*)+ 0.09845 \; {\cal I} (\delta R_{9})+  2.73 \; {\cal R} (\delta R_{10})+ 0.001146 \; {\cal R} ( \delta R_{10} \delta R_7^*)
\nonumber\\ & & 
+ 0.0001236 \; {\cal R} ( \delta R_{10} \delta R_8^*)- 0.0173 \; {\cal R} ( \delta R_{10} \delta R_9^*)- 0.1119 \; {\cal R} (\delta R_{7})- 0.01279 \; {\cal R} (\delta R_{8})
\nonumber\\ & & 
+ 0.0006912 \; {\cal R} ( \delta R_8 \delta R_7^*)+ 0.8243 \; {\cal R} (\delta R_{9})- 0.04872 \; {\cal R} ( \delta R_9 \delta R_7^*)
\nonumber\\ & & 
- 0.005478 \; {\cal R} ( \delta R_9 \delta R_8^*)+ 1.411 \; |\delta R_{10}|^2+ 0.002655 \; |\delta R_7|^2+ 0.00003702 \; |\delta R_8|^2
\nonumber\\ & & 
+ 0.2016 \; |\delta R_9|^2
\; \Big] \times 10^{-7} \; ,
\eea
where  
\bea
1+\delta R_{7,8} = \frac{C_{7,8}^{(00){\rm eff}} (\mu_0)}{C_{7,8}^{(00){\rm eff,SM}}(\mu_0)}  
\hspace{1cm} {\rm and} \hspace{1cm}
1+\delta R_{9,10} = \frac{C_{9,10}^{(11)} (\mu_0)}{C_{9,10}^{(11){\rm SM}}(\mu_0)} \;,
\eea
and we refer the reader to the definition of the Wilson coefficients given in Ref.~\cite{Huber:2005ig}.

The analogous formulae for the integrated forward backward 
asymmetry in bin 1 and bin 2 are (we give separately the FBA from Eq.~(\ref{fba}), ${\cal A}^{FB}_{\mu\mu}$, and the normalizing branching ratio from Eq.~(\ref{br}), ${\cal B}_{\mu\mu}$):
\bea
\left({\cal A}_{\mu\mu}^{FB}\right)_{\rm bin 1} & = & \Big[\;
-7.877-  0.84 \; {\cal I} (\delta R_{10})+ 0.1901 \; {\cal I} 
(\delta R_{8})+ 0.1901 \; {\cal I} ( \delta R_8 \delta R_{10}^*)
\nonumber\\ & &
+ 0.02323 \; {\cal I} (\delta R_{9})- 8.136 \; {\cal R} (\delta R_
{10})+ 0.02323 \; {\cal I} (\delta R_{9}) \; {\cal R} (\delta R_{10})
\nonumber\\ & &
- 6.114 \; {\cal R} ( \delta R_{10} \delta R_7^*)- 0.6183 \; {\cal 
R} ( \delta R_{10} \delta R_8^*)+ 5.048 \; {\cal R} ( \delta R_{10} 
\delta R_9^*)
\nonumber\\ & &
- 5.942 \; {\cal R} (\delta R_{7})- 0.6059 \; {\cal R} (\delta R_
{8})+ 4.947 \; {\cal R} (\delta R_{9})- 0.02323 \; {\cal I} (\delta 
R_{10}) \; {\cal R} (\delta R_{9})
\nonumber\\ & &
+ 0.01569 \; {\cal R} ( \delta R_9 \delta R_7^*)- 0.08693 \; |
\delta R_{10}|^2- 0.0128 \; |\delta R_9|^2
\;\Big] \times 10^{-8} \; ,
\eea
\bea
\left({\cal B}_{\mu\mu}\right)_{\rm bin 1} & = & \Big[\;
8.653+ 0.07321 \; {\cal I} (\delta R_{7})+ 0.0167 \; {\cal I} 
( \delta R_7 \delta R_8^*)+ 0.002753 \; {\cal I} ( \delta R_7 
\delta R_9^*)
\nonumber\\ & &
- 0.003923 \; {\cal I} (\delta R_{8})+ 0.01471 \; {\cal I} 
( \delta R_8 \delta R_9^*)- 0.03578 \; {\cal I} (\delta R_{9})+ 
10.62 \; {\cal R} (\delta R_{10})
\nonumber\\ & &
+ 0.008387 \; {\cal R} ( \delta R_{10} \delta R_7^*)+ 0.001049 \; 
{\cal R} ( \delta R_{10} \delta R_8^*)- 0.05413 \; {\cal R} 
( \delta R_{10} \delta R_9^*)
\nonumber\\ & &
+ 0.262 \; {\cal R} (\delta R_{7})+ 0.02037 \; {\cal R} (\delta R_
{8})+ 0.05047 \; {\cal R} ( \delta R_8 \delta R_7^*)+ 2.264 \; 
{\cal R} (\delta R_{9})
\nonumber\\ & &
- 0.4778 \; {\cal R} ( \delta R_9 \delta R_7^*)- 0.05645 \; {\cal 
R} ( \delta R_9 \delta R_8^*)+ 5.463 \; |\delta R_{10}|^2+ 0.2087 
\; |\delta R_7|^2
\nonumber\\ & &
+ 0.002892 \; |\delta R_8|^2+ 0.7634 \; |\delta R_9|^2
\;\Big] \times 10^{-7} \; ,
\eea
\bea
\left({\cal A}_{\mu\mu}^{FB}\right)_{\rm bin 2} & = & \Big[\;
5.459- 0.7026 \; {\cal I} (\delta R_{10})+ 0.1426 \; {\cal I} 
(\delta R_{8})+ 0.1426 \; {\cal I} ( \delta R_8 \delta R_{10}^*)
\nonumber\\ & &
+ 0.03706 \; {\cal I} (\delta R_{9})+ 5.472 \; {\cal R} (\delta R_
{10})+ 0.03706 \; {\cal I} (\delta R_{9}) \; {\cal R} (\delta R_{10})
\nonumber\\ & &
-  4.48 \; {\cal R} ( \delta R_{10} \delta R_7^*)- 0.4645 \; {\cal 
R} ( \delta R_{10} \delta R_8^*)+ 7.832 \; {\cal R} ( \delta R_{10} 
\delta R_9^*)
\nonumber\\ & &
- 4.352 \; {\cal R} (\delta R_{7})- 0.4645 \; {\cal R} (\delta R_
{8})+ 7.622 \; {\cal R} (\delta R_{9})- 0.03706 \; {\cal I} (\delta 
R_{10}) \; {\cal R} (\delta R_{9})
\nonumber\\ & &
+ 0.01172 \; {\cal R} ( \delta R_9 \delta R_7^*)- 0.1382 \; |\delta 
R_{10}|^2- 0.02034 \; |\delta R_9|^2
\;\Big] \times 10^{-8} \; ,
\eea
\bea
\left({\cal B}_{\mu\mu}\right)_{\rm bin 2} & = & \Big[\;
7.052+ 0.02856 \; {\cal I} (\delta R_{7})+ 0.005661 \; {\cal I} 
( \delta R_7 \delta R_8^*)+ 0.002082 \; {\cal I} ( \delta R_7 
\delta R_9^*)
\nonumber\\ & &
+ 0.01401 \; {\cal I} (\delta R_{8})+ 0.01112 \; {\cal I} ( \delta 
R_8 \delta R_9^*)- 0.03512 \; {\cal I} (\delta R_{9})+ 9.472 \; 
{\cal R} (\delta R_{10})
\nonumber\\ & &
+ 0.006338 \; {\cal R} ( \delta R_{10} \delta R_7^*)- 0.04942 \; 
{\cal R} ( \delta R_{10} \delta R_9^*)- 0.4596 \; {\cal R} (\delta 
R_{7})
\nonumber\\ & &
- 0.05505 \; {\cal R} (\delta R_{8})+ 0.01678 \; {\cal R} ( \delta 
R_8 \delta R_7^*)+ 2.394 \; {\cal R} (\delta R_{9})
\nonumber\\ & &
- 0.3527 \; {\cal R} ( \delta R_9 \delta R_7^*)
- 0.04016 \; {\cal R} ( \delta R_9 \delta R_8^*)+ 4.879 \; |\delta 
R_{10}|^2
\nonumber\\ & &
+ 0.07016 \; |\delta R_7|^2+ 0.6891 \; |\delta R_9|^2
\;\Big] \times 10^{-7} \; .
\eea
The integrated asymmetries in bin 1, bin 2, and in the whole low-$q^2$ 
region are then according to Eq.~(\ref{doubleratio}):
\bea
\left( \bar{\cal A}_{\mu\mu}^{FB}\right)_{\rm bin 1} & = & \frac{\left({\cal 
A}_{\mu\mu}^{FB}\right)_{\rm bin 1}}{
\left({\cal B}_{\mu\mu}\right)_{\rm bin 1}} \; , \\
\left( \bar{\cal A}_{\mu\mu}^{FB}\right)_{\rm bin 2} & = & \frac{\left({\cal 
A}_{\mu\mu}^{FB}\right)_{\rm bin 2}}{
\left({\cal B}_{\mu\mu}\right)_{\rm bin 2}} \; , \\
\left( \bar{\cal A}_{\mu\mu}^{FB}\right)_{\rm low} & = & \frac{
\left({\cal A}_{\mu\mu}^{FB}\right)_{\rm bin 1}+\left({\cal A}_{\mu
\mu}^{FB}\right)_{\rm bin 2}}{
\left({\cal B}_{\mu\mu}\right)_{\rm bin 1}+\left({\cal B}_{\mu\mu}
\right)_{\rm bin 2}} \; .
\eea

\section{Summary and Outlook}
\label{sec:conclusions}
In this paper we extend the analysis of log--enhanced QED corrections initiated in Ref.~\cite{Huber:2005ig} to the $\bar B \to X_s \ell
\ell$ decay width in the high-$q^2$ region and to the forward backward asymmetry.

We give a complete phenomenological analysis of all relevant quantities related to these observables, including the ratio ${\cal
R}(s_0)$ recently proposed in Refs.~\cite{Ligeti:2007sn,Bauer:2000xf}. We propose a new approach to
the extraction of the zero of the FBA and argue that the scale dependence obtained by this procedure is a reasonable reflection of the
perturbative error.

It is well-known that the measurements of those quantities, in addition to the $\bar B \rightarrow X_s \gamma$ branching ratio, will allow to fix magnitude
and sign of all relevant Wilson coefficients in the SM. Since at the end of the current $B$ factories 
quantities integrated over certain $q^2$-bins will be already accessible with high precision,
the following observation is important, namely  that the double differential decay width decomposed 
according to ($z =\cos\theta$)
\be
\frac{d^2\Gamma}{dq^2 \, dz} = \frac{3}{8} \, \left[(1+z^2) \, H_T(q^2) \, + \, 2 \, z \, H_A(q^2) +2 \, (1-z^2) \, H_L(q^2)\right] \; ,
\ee
where
\be
\frac{d\Gamma}{dq^2} = H_T(q^2) \, +H_L(q^2)\, , \qquad \frac{dA_{\rm{FB}}}{dq^2} = 3/4 \, H_A(q^2) \, ,
\ee
includes a third quantity which depends on a different combination of Wilson coefficients~\cite{Lee:2006gs}. The NNLO+QED analysis of this observable will be published in a forthcoming paper.

\section*{Acknowledgments}
We would like to thank Martin Beneke,  Christoph Greub, Ulrich Haisch, 
Gino Isidori, Miko{\l}aj Misiak, Lalit Sehgal, and Peter Uwer for
interesting and helpful discussions and suggestions. This work was 
supported by Schweizerischer Nationalfonds
(SNF) and by Deutsche Forschungsgemeinschaft, SFB/TR 9 ``Computergest\"{u}tzte Theoretische Teilchenphysik''. Fermilab is operated by
Fermi Research Alliance, LLC under Contract No. DE-AC02-07CH11359 with the United States Department
of Energy.

\setlength {\baselineskip}{0.2in}


\begin{thebibliography}{99}

\newcommand{\np}[3]{Nucl. Phys. {\bf B#1} (#2) #3}
\newcommand{\pl}[3]{Phys. Lett. {\bf B#1} (#2) #3}
\newcommand{\pr}[3]{Phys. Rev.  {\bf D#1} (#2) #3}
\newcommand{\prl}[3]{Phys. Rev. Lett. {\bf #1} (#2) #3}
\newcommand{\prp}[3]{Phys. Rept. {\bf #1} (#2) #3}
\newcommand{\ptp}[3]{Prog. Theor. Phys. {\bf #1} (#2) #3}
\newcommand{\zpc}[3]{Z. Phys. {\bf C#1} (#2) #3}
\newcommand{\ibid}[3]{{\it ibid.} {\bf #1} (#2) #3}

\bibitem{Hurth:2007xa}
  T.~Hurth,
  %``Status of SM calculations of b > s transitions,''
  Int.\ J.\ Mod.\ Phys.\  A {\bf 22} (2007) 1781
  [arXiv:hep-ph/0703226].
  %%CITATION = IMPAE,A22,1781;%%

\bibitem{Hurth:2003vb}
  T.~Hurth,
  %``Present status of inclusive rare B decays,''
  Rev.\ Mod.\ Phys.\  {\bf 75} (2003) 1159
  [arXiv:hep-ph/0212304].
  %%CITATION = RMPHA,75,1159;%%

\bibitem{Iwasaki:2005sy}
  M.~Iwasaki {\it et al.}  [Belle Collaboration],
  %``Improved measurement of the electroweak penguin process B $\to$ X/s l+l-,''
  hep-ex/0503044.
  %%CITATION = HEP-EX 0503044;%%

\bibitem{Aubert:2004it}
B.~Aubert {\it et al.}  [BABAR Collaboration],
%``Measurement of the B $\to$ X/s l+ l- branching fraction with a sum over exclusive modes,''
Phys.\ Rev.\ Lett.\  {\bf 93} (2004) 081802
[hep-ex/0404006].
%%CITATION = HEP-EX 0404006;%%

\bibitem{MISIAKBOBETH}
  C.~Bobeth, M.~Misiak and J.~Urban,
  %``Photonic penguins at two loops and m(t)-dependence of BR(B $\to$ X(s) l+
  %l-),''
  Nucl.\ Phys.\ B {\bf 574} (2000) 291
  [hep-ph/9910220].
  %%CITATION = HEP-PH 9910220;%%

\bibitem{Asa1}
  H.~H.~Asatryan, H.~M.~Asatrian, C.~Greub and M.~Walker,
  %``Calculation of two loop virtual corrections to b $\to$ s l+ l- in the
  %standard model,''
  Phys.\ Rev.\ D {\bf 65} (2002) 074004
  [hep-ph/0109140].
  %%CITATION = HEP-PH 0109140;%%

\bibitem{Asatryan:2002iy}
H.~H.~Asatryan, H.~M.~Asatrian, C.~Greub and M.~Walker,
%``Complete gluon bremsstrahlung corrections to the process b $\to$ s l+ l-,''
Phys.\ Rev.\ D {\bf 66} (2002) 034009
[hep-ph/0204341].
%%CITATION = HEP-PH 0204341;%%

\bibitem{Adrian2}
  A.~Ghinculov, T.~Hurth, G.~Isidori and Y.~P.~Yao,
  %``Forward-backward asymmetry in B $\to$ X/s l+ l- at the NNLL level,''
  Nucl.\ Phys.\ B {\bf 648} (2003) 254
  [hep-ph/0208088].
  %%CITATION = HEP-PH 0208088;%%

\bibitem{Asatrian:2002va}
  H.~M.~Asatrian, K.~Bieri, C.~Greub and A.~Hovhannisyan,
  %``NNLL corrections to the angular distribution and to the forward-backward
  %asymmetries in b $\to$ X/s l+ l-,''
  Phys.\ Rev.\ D {\bf 66} (2002) 094013
  [hep-ph/0209006].
  %%CITATION = HEP-PH 0209006;%%

\bibitem{Ghinculov:2003bx}
  A.~Ghinculov, T.~Hurth, G.~Isidori and Y.~P.~Yao,
  %``New NNLL QCD results on the decay B --> X/s l+ l-,''
  Eur.\ Phys.\ J.\  C {\bf 33} (2004) S288
  [arXiv:hep-ph/0310187].
  %%CITATION = EPHJA,C33,S288;%%

\bibitem{Adrian1}
  A.~Ghinculov, T.~Hurth, G.~Isidori and Y.~P.~Yao,
  %``The rare decay B $\to$ X/s l+ l- to NNLL precision for arbitrary dilepton
  %invariant mass,''
  Nucl.\ Phys.\ B {\bf 685} (2004) 351
  [hep-ph/0312128].
  %%CITATION = HEP-PH 0312128;%%

\bibitem{Gambinonew} 
  C.~Bobeth, P.~Gambino, M.~Gorbahn and U.~Haisch,
  %``Complete NNLO QCD analysis of anti-B $\to$ X/s l+ l- and higher order
  %electroweak effects,''
  JHEP {\bf 0404}, 071 (2004)
  [hep-ph/0312090].
  %%CITATION = HEP-PH 0312090;%%

\bibitem{Asatrian:2003yk}
  H.~M.~Asatrian, H.~H.~Asatryan, A.~Hovhannisyan and V.~Poghosyan,
  %``Complete bremsstrahlung corrections to the forward-backward asymmetries  in
  %b --> X/s l+ l-,''
  Mod.\ Phys.\ Lett.\  A {\bf 19} (2004) 603
  [arXiv:hep-ph/0311187].
  %%CITATION = MPLAE,A19,603;%%

\bibitem{Gorbahn:2004my}
M.~Gorbahn and U.~Haisch,
%``Effective Hamiltonian for non-leptonic $|$Delta(F)$|$ = 1 decays
%at NNLO in QCD,''
Nucl.\ Phys.\ B {\bf 713} (2005) 291
[hep-ph/0411071].
%%CITATION = HEP-PH 0411071;%%

\bibitem{Akeroyd:2004mj}
  A.~G.~Akeroyd {et al.},
  %``Physics at super B factory,''
  arXiv:hep-ex/0406071.
  %%CITATION = HEP-EX 0406071;%%
 
\bibitem{Hewett:2004tv}
  J.~Hewett {et al.},   arXiv:hep-ph/0503261.
  %%CITATION = HEP-PH 0503261;%%

\bibitem{Bona:2007qt}
  M.~Bona {\it et al.},
  %``SuperB: A High-Luminosity Asymmetric e+ e- Super Flavor Factory. Conceptual
  %Design Report,''
  arXiv:0709.0451 [hep-ex].
  %%CITATION = ARXIV:0709.0451;

\bibitem{Falk}
  A.~F.~Falk, M.~E.~Luke and M.~J.~Savage,
  % ``Nonperturbative contributions to the inclusive rare decays B $\to$ X(s)
  %gamma and B $\to$ X(s) lepton+ lepton-,''
  Phys.\ Rev.\ D {\bf 49}, 3367 (1994)
  [arXiv:hep-ph/9308288].
  %%CITATION = HEP-PH 9308288;%%

\bibitem{Alineu}
A.~Ali, G.~Hiller, L.~T.~Handoko and T.~Morozumi,
%``Power corrections in the decay rate and distributions 
%in  $B \to X_s \ell^+ \ell^-$ in the standard model,''
Phys.\ Rev.\ D {\bf 55}, 4105 (1997) [hep-ph/9609449].

\bibitem{Savagenew}
J.~W.~Chen, G.~Rupak and M.~J.~Savage,
%``Non-$1/m_b^n$ power suppressed contributions to inclusive  $B \to X_s
%\ell^+ \ell^-$ decays,''
Phys.\ Lett.\ B {\bf 410}, 285 (1997) [hep-ph/9705219].

\bibitem{Buchalla:1997ky}
  G.~Buchalla, G.~Isidori and S.~J.~Rey,
  %``Corrections of order Lambda(QCD)**2/m(c)**2 to inclusive rare B  decays,''
  Nucl.\ Phys.\ B {\bf 511}, 594 (1998)
  [arXiv:hep-ph/9705253].
  %%CITATION = HEP-PH 9705253;%%

\bibitem{buchallanewnew} 
G.~Buchalla and G.~Isidori,
%``Non-perturbative effects in $\bar B \to X_s \ell^+ \ell^-$ 
%for large dilepton  invariant mass,''
Nucl.\ Phys.\ B {\bf 525}, 333 (1998) [hep-ph/9801456].

\bibitem{Bauer}
C.~W.~Bauer and C.~N.~Burrell,
%``Non-perturbative corrections to moments of the decay 
%$B \to X_s \ell^+ \ell^-$,''
Phys.\ Rev.\ D {\bf 62}, 114028 (2000) [hep-ph/9911404].

\bibitem{Ligeti:2007sn}
  Z.~Ligeti and F.~J.~Tackmann,
  %``Precise predictions for B -> Xs l+ l- in the large q^2 region,''
  Phys.\ Lett.\  B {\bf 653} (2007) 404
  [arXiv:0707.1694v2].
  
\bibitem{KS}
  F.~Kruger and L.~M.~Sehgal,
  %``Lepton Polarization in the Decays $B\to X_s\mu~+\mu~-$ and $B\to
  %X_s\tau~+\tau~-$,''
  Phys.\ Lett.\  B {\bf 380} (1996) 199
  [arXiv:hep-ph/9603237].  


\bibitem{Neubert}
%\cite{Neubert:2000ch}
%\bibitem{Neubert:2000ch}
  M.~Neubert,
  %``On the inclusive determination of |V(ub)| from the lepton invariant  mass
  %spectrum,''
  JHEP {\bf 0007} (2000) 022
  [arXiv:hep-ph/0006068].
  %%CITATION = JHEPA,0007,022;%%




\bibitem{BLK} 
%\cite{Bauer:2001rc}
%\bibitem{Bauer:2001rc}
  C.~W.~Bauer, Z.~Ligeti and M.~E.~Luke,
  %``Precision determination of |V(ub)| from inclusive decays,''
  Phys.\ Rev.\  D {\bf 64} (2001) 113004
  [arXiv:hep-ph/0107074].
  %%CITATION = PHRVA,D64,113004;%%




\bibitem{Lee:2005pk}
  K.~S.~M.~Lee and I.~W.~Stewart,
  %``Shape-function effects and split matching in B --> X/s l+ l-,''
  Phys.\ Rev.\ D {\bf 74} (2006) 014005
  [hep-ph/0511334].
  %%CITATION = HEP-PH 0511334;%%

\bibitem{Lee:2005pw}
  K.~S.~M.~Lee, Z.~Ligeti, I.~W.~Stewart and F.~J.~Tackmann,
  %``Universality and m(X) cut effects in B --> X/s l+ l-,''
  Phys.\ Rev.\ D {\bf 74} (2006) 011501
  [hep-ph/0512191].
  %%CITATION = HEP-PH 0512191;%%

\bibitem{Huber:2005ig}
  T.~Huber, E.~Lunghi, M.~Misiak and D.~Wyler,
  %``Electromagnetic logarithms in anti-B --> X/s l+ l-,''
  Nucl.\ Phys.\ B {\bf 740} (2006) 105
  [hep-ph/0512066].
  %%CITATION = HEP-PH 0512066;%%

\bibitem{Bauer:2004ve}
  C.~W.~Bauer, Z.~Ligeti, M.~Luke, A.~V.~Manohar and M.~Trott,
  %``Global analysis of inclusive B decays,''
  Phys.\ Rev.\ D {\bf 70}, 094017 (2004)
  [hep-ph/0408002].
  %%CITATION = HEP-PH 0408002;%%

\bibitem{semi3}
  I.~Blokland, A.~Czarnecki, M.~Slusarczyk and F.~Tkachov,
  %``Next-to-next-to-leading order calculations for heavy-to-light decays,''
  Phys.\ Rev.\  D {\bf 71} (2005) 054004
  [arXiv:hep-ph/0503039].
  %%CITATION = PHRVA,D71,054004;%%

\bibitem{Voloshin:2001xi}
  M.~B.~Voloshin,
  %``Nonfactorization effects in heavy mesons and determination of |V(ub)|  from
  %inclusive semileptonic B decays,''
  Phys.\ Lett.\  B {\bf 515}, 74 (2001)
  [arXiv:hep-ph/0106040].
  %%CITATION = PHLTA,B515,74;%%

\bibitem{Ali:1991is}
  A.~Ali, T.~Mannel and T.~Morozumi,
  %``Forward backward asymmetry of dilepton angular distribution in the decay b
  %$\to$ s l+ l-,''
  Phys.\ Lett.\  B {\bf 273} (1991) 505.
  %%CITATION = PHLTA,B273,505;%%

\bibitem{Hoang:1998hm}
   A.~H.~Hoang, Z.~Ligeti and A.~V.~Manohar,
  %``B decays in the Upsilon expansion,''
  Phys.\ Rev.\ D {\bf 59} (1999) 074017
  [hep-ph/9811239]. 
  %%CITATION = HEP-PH 9811239;%%

\bibitem{Hoang:2000fm}
  A.~H.~Hoang,
  %``Bottom quark mass from Upsilon mesons: Charm mass effects,''
  hep-ph/0008102.
  %%CITATION = HEP-PH 0008102;%%

\bibitem{semi1}
  A.~Czarnecki and K.~Melnikov,
  %``Semileptonic b --> u decays: Lepton invariant mass spectrum,''
  Phys.\ Rev.\ Lett.\  {\bf 88}, 131801 (2002)
  [arXiv:hep-ph/0112264].
  %%CITATION = PRLTA,88,131801;%%

\bibitem{semi2}
  K.~G.~Chetyrkin, R.~Harlander, T.~Seidensticker and M.~Steinhauser,
  %``Second order {QCD} corrections to Gamma(t --> W b),''
  Phys.\ Rev.\  D {\bf 60} (1999) 114015
  [arXiv:hep-ph/9906273].
  %%CITATION = PHRVA,D60,114015;%%

\bibitem{semi4}
  I.~Blokland, A.~Czarnecki, M.~Slusarczyk and F.~Tkachov,
  %``Heavy-to-light decays with a two-loop accuracy,''
  Phys.\ Rev.\ Lett.\  {\bf 93}, 062001 (2004)
  [arXiv:hep-ph/0403221].
  %%CITATION = PRLTA,93,062001;%

\bibitem{Bethke:2006ac}
  S.~Bethke,
  %``Experimental tests of asymptotic freedom,''
  Prog.\ Part.\ Nucl.\ Phys.\  {\bf 58}, 351 (2007)
  [arXiv:hep-ex/0606035].
  %%CITATION = PPNPD,58,351;%%
  
\bibitem{Charles:2004jd}
  J.~Charles {\it et al.}  [CKMfitter Group],
  %``CP violation and the CKM matrix: Assessing the impact of the asymmetric  B
  %factories,''
  Eur.\ Phys.\ J.\ C {\bf 41}, 1 (2005)
  [hep-ph/0406184].
  %%CITATION = HEP-PH 0406184;%%

\bibitem{Hoang:2005zw}
   A.~H.~Hoang and A.~V.~Manohar,
  %``Charm quark mass from inclusive semileptonic B decays,''
  hep-ph/0509195. 
  %%CITATION = HEP-PH 0509195;%%

\bibitem{Aubert:2004aw}
B.~Aubert {\it et al.}  (BaBar Collaboration),
%``Determination of the branching fraction for B $\to$ X/c l nu decays and of
%$|$V(cb)$|$ from hadronic mass and lepton energy moments,''
Phys.\ Rev.\ Lett.\  {\bf 93}, 011803 (2004) [hep-ex/0404017]. 
%%CITATION = HEP-EX 0404017;%%

\bibitem{topmass}
  P.~A.~Movilla Fernandez  [CDF Collaboration],
  %``Precision Determination of the Top Quark Mass,''
  arXiv:0705.3910 [hep-ex].
  %%CITATION = ARXIV:0705.3910;%%
  
\bibitem{Eidelman:2004wy}
  S.~Eidelman {\it et al.}  [Particle Data Group],
  %``Review of particle physics,''
  Phys.\ Lett.\ B {\bf 592}, 1 (2004).
  %%CITATION = PHLTA,B592,1;%%

\bibitem{private}  J.~Berryhill, A.~Ishikawa, private communication. 
  
\bibitem{AGHL}
  A.~Ali, E.~Lunghi, C.~Greub and G.~Hiller,
  %``Improved model-independent analysis of semileptonic and radiative rare  B
  %decays,''
  Phys.\ Rev.\  D {\bf 66} (2002) 034002
  [arXiv:hep-ph/0112300].
  
  %%CITATION = PHRVA,D66,034002;%%
\bibitem{Boughezal:2006px}
  R.~Boughezal, M.~Czakon and T.~Schutzmeier,
  %``Charm and bottom quark masses from perturbative QCD,''
  Phys.\ Rev.\  D {\bf 74} (2006) 074006
  [arXiv:hep-ph/0605023].
  %%CITATION = PHRVA,D74,074006;%%

\bibitem{Bauer:2000xf}
  C.~W.~Bauer, Z.~Ligeti and M.~E.~Luke,
  %``A model independent determination of |V(ub)|,''
  Phys.\ Lett.\  B {\bf 479} (2000) 395
  [arXiv:hep-ph/0002161].
  %%CITATION = PHLTA,B479,395;%%

\bibitem{Lee:2006gs}
  K.~S.~M.~Lee, Z.~Ligeti, I.~W.~Stewart and F.~J.~Tackmann,
  %``Extracting short distance information from b --> s l+ l- effectively,''
  Phys.\ Rev.\  D {\bf 75} (2007) 034016
  [arXiv:hep-ph/0612156].
  %%CITATION = PHRVA,D75,034016;%%

\end{thebibliography}
\end{document}